\documentclass[12pt]{article}
\usepackage[latin1]{inputenc}
\usepackage[dvips]{graphicx,color}
\usepackage{amsmath}
\usepackage{amsfonts}
\usepackage{amssymb}
\usepackage{graphicx}
\usepackage{geometry}
\usepackage{amssymb,epsfig,subfigure}
\usepackage{graphicx}

\makeatletter
\renewcommand\section{\@startsection {section}{1}{\z@}%
                                 {-3.5ex \@plus -1ex \@minus -.2ex}
                                   {2.3ex \@plus.2ex}%
                                   {\normalfont\large\bfseries}}
\renewcommand\subsection{\@startsection{subsection}{2}{\z@}%
                                   {-3.25ex\@plus -1ex \@minus -.2ex}%
                                     {1.5ex \@plus .2ex}%
                                     {\normalfont\bfseries}}
\renewcommand\subsubsection{\@startsection{subsubsection}{3}{\z@}%
                                   {-3.25ex\@plus -1ex \@minus -.2ex}%
                                     {1.5ex \@plus .2ex}%
                                     {\normalfont\itshape}}
\makeatother

\def\pplogo{\vbox{\kern-\headheight\kern -29pt
\halign{##&##\hfil\cr&{\ppnumber}\cr\rule{0pt}{2.5ex}&\ppdate\cr}}}
\makeatletter
\def\ps@firstpage{\ps@empty \def\@oddhead{\hss\pplogo}%
  \let\@evenhead\@oddhead 
}
\def\maketitle{\par
 \begingroup
 \def\thefootnote{\fnsymbol{footnote}}
 \def\@makefnmark{\hbox{$^{\@thefnmark}$\hss}}
 \if@twocolumn
 \twocolumn[\@maketitle]
 \else \newpage
 \global\@topnum\z@ \@maketitle \fi\thispagestyle{firstpage}\@thanks
 \endgroup
 \setcounter{footnote}{0}
 \let\maketitle\relax
 \let\@maketitle\relax
 \gdef\@thanks{}\gdef\@author{}\gdef\@title{}\let\thanks\relax}
\makeatother

\numberwithin{equation}{section}

\renewcommand{\dag}{\dagger}

\newcommand{\be}{\begin{eqnarray}}
\newcommand{\bea}{\begin{eqnarray}}
\newcommand{\ee}{\end{eqnarray}}
\newcommand{\eea}{\end{eqnarray}}

\newcommand{\f}{\frac}
\newcommand{\mc}{\mathcal}
\newcommand{\Tr}{{\rm Tr}}
\newcommand{\tr}{{\rm tr}}
\renewcommand{\t}{\tilde}
\newcommand{\muphi}{\mu_\phi}

\begin{document}

\setcounter{page}0
\def\ppnumber{\vbox{\baselineskip14pt
}}
\def\ppdate{\footnotesize{\tt SLAC-PUB-14088, 
NSF-KITP-10-054}} \date{}

\author{Sakura Sch\"afer-Nameki$^{1}$, Carlos Tamarit$^{1}$ and Gonzalo Torroba$^{2\,,1}$\\
[7mm]
{\normalsize $^1$ \it Kavli Institute for Theoretical Physics}\\
{\normalsize \it University of California, Santa Barbara, CA 93106, USA}\\
{\normalsize $^2$ \it SLAC National Accelerator Laboratory, }\\
{\normalsize \it Stanford University, Stanford, CA 94309, USA}\\
[3mm]
{\tt \footnotesize ss299 at  kitp.ucsb.edu, tamarit at kitp.ucsb.edu, torrobag at slac.stanford.edu }
}

\bigskip
\title{\bf  A Hybrid Higgs\\
\vskip 0.5cm}
\maketitle

\begin{abstract} \normalsize
\noindent
We construct composite Higgs models admitting a weakly coupled Seiberg dual description. We focus on the possibility that only the up-type Higgs is an elementary field, while the down-type Higgs arises as a composite hadron. The model, based on a confining SQCD theory, breaks supersymmetry and electroweak symmetry dynamically and calculably. This simultaneously solves the $\mu/B_\mu$ problem and explains the smallness of the bottom and tau masses compared to the top mass. The proposal is then applied to a class of models where the same confining dynamics is used to generate the Standard Model flavor hierarchy by quark and lepton compositeness. This provides a unified framework for flavor, supersymmetry breaking and electroweak physics. The weakly coupled dual is used to explicitly compute the MSSM parameters in terms of a few microscopic couplings, giving interesting relations between the electroweak and soft parameters. The RG evolution down to the TeV scale is obtained and salient phenomenological predictions of this class of ``single-sector'' models are discussed. 
\end{abstract}
\bigskip
\newpage

\tableofcontents

\vskip 1cm

\section{Introduction}\label{sec:intro}

The Standard Model (SM) has deep theoretical puzzles that hint toward the existence of a more fundamental microscopic theory. First, the SM fermions exhibit very special patterns of Yukawa couplings; the origin of these ``flavor spurions'' is unknown. Another central question concerns the hierarchy between the Fermi and Planck scales. Supersymmetry offers a very attractive mechanism for stabilizing $M_Z$, but does not explain its origin. Fits to precision electroweak data suggest a weakly coupled Higgs and in the SM its mass is a free parameter whose origin is unknown. There are various solutions to each of these problems separately, but the different mechanisms are in general hard to combine. Nevertheless, trying to find a single unified framework addressing these puzzles can reveal new connections between flavor, supersymmetry breaking and Higgs physics, as well as providing windows into high energy physics.

Based on~\cite{ArkaniHamed:1997fq, Luty:1998vr, Franco:2009wf}, the authors of~\cite{Craig:2009hf} proposed a realistic ``single-sector'' model where the dynamics that explains the texture of fermion masses also breaks supersymmetry.\footnote{The original models of~\cite{ArkaniHamed:1997fq, Luty:1998vr} contained strongly coupled incalculable effects; in~\cite{Franco:2009wf} it was understood how to construct calculable single sector models.} The confining dynamics is given by supersymmetric QCD with fundamental flavors plus a field in the adjoint representation, in the free magnetic phase. The hierarchies in the Yukawa matrices are explained by postulating that the first and second SM generations are composite mesons of different UV dimensions, while the third generation is elementary.\footnote{SUSY models using compositeness to explain the flavor hierarchies were constructed in~\cite{Strassler:1995ia}; related ideas make use of conformal dynamics~\cite{NS, poland, Aharony:2010ch,Craig:2010ip}. AdS/CFT can also be used to understand the generation of flavor hierarchies at strong coupling; see e.g.~\cite{Gabella:2007cp}. We refer the reader to~\cite{Craig:2009hf} for a recent overview of approaches to the flavor problem.}
Supersymmetry is broken dynamically using a variant of the mechanism found by Intriligator, Seiberg and Shih (ISS)~\cite{Intriligator:2006dd}. The model is fully calculable and realistic, avoids flavor problems and produces a low energy spectrum similar to the ``more minimal'' supersymmetric SM of~\cite{Cohen:1996vb,Dimopoulos:1995mi}.

The original motivation of this work was to study the Higgs sector that could arise in the single sector models of~\cite{Craig:2009hf}. However, the dynamical mechanism that we found turns out to be quite generic and simple, and can be applied to theories with a flavor sector different from the one in~\cite{Craig:2009hf}. We will argue that SQCD with fundamental flavors in the free magnetic phase has the required structure to produce a composite Higgs model that breaks the electroweak symmetry dynamically and calculably, and naturally solves the $\mu/B_\mu$ problem.\footnote{See~\cite{Komargodski:2008ax} for a recent analysis and references for this problem in the context of gauge mediation.}

Therefore, the first part of the paper (\S \ref{sec:compositeH} and \S\ref{sec:dewsb}) will be devoted to presenting the mechanism for dynamical EWSB in its simplest and minimal form, independently of the flavor sector. The precise low energy phenomenology at the TeV scale depends on the spectrum and couplings to the (supersymmetric) SM particles, a well-known example being the sensitivity of the Higgs to top/stop effects. In the second part of the work (starting from \S \ref{sec:ssector}), we will show how our proposal naturally combines with the single sector models of~\cite{Craig:2009hf}, yielding a unified and realistic explanation for flavor, supersymmetry breaking and Higgs physics.

The model has various interesting properties, stemming from the fact that the strong dynamics responsible for the breaking of $SU(2) \times U(1)$ also produces the fermion masses and mixings and breaks supersymmetry. For instance, the EW scale is tied to the supersymmetry scale via a coupling which is also responsible for producing the Yukawa interactions. The $\mu$ term is related to the scale of R-symmetry breaking and gaugino masses. The RG evolution of the theory down to the TeV scale presents various special features, studied in \S \ref{sec:eft}. \S \ref{sec:pheno} presents the low energy phenomenology and a scan over parameter space, with particular emphasis on the nature of the NLSP. Before proceeding, it is useful to summarize the basic ideas and present an overview of the model.

\subsection{The basic mechanism}\label{subsec:calctechni}

Strongly coupled gauge theories have a very elegant mechanism for generating exponentially small hierarchies,
$$
\Lambda_{IR} \sim e^{-g_{IR}^2/g_{UV}^2}\, \Lambda_{UV}\,.
$$
Both dynamical supersymmetry breaking and technicolor exploit this fact. In the latter case, the electroweak scale is generated by identifying the Higgs field with a technifermion bilinear $H \equiv \bar \psi \psi$ which condenses due to nonperturbative effects from a technicolor gauge group $G_{TC}$~\cite{technicolor}. 

Technicolor constructions generically face the problem of new strongly coupled dynamics close to the TeV scale. Some of the difficulties are overcome if the physical dimension of $H$ is close to one (e.g. walking technicolor), although this limit is also somewhat problematic. Unitarity implies that $H$ must become free when its dimension is equal to one, but in 4d it is challenging to find a realistic technicolor theory supporting weakly coupled subsectors. Furthermore, the operator $H^* H$ becomes relevant and the hierarchy and fine-tuning problems reappear. 

These points can in principle be resolved in a supersymmetric context. Weakly coupled subsectors are quite common in supersymmetric confining gauge theories -- the simplest being SQCD in the free magnetic phase. And, of course, one of the main motivations for supersymmetry is that it stabilizes the Fermi scale, making the theory natural up to extremely high scales. Our goal (in \S \ref{sec:compositeH} and \S \ref{sec:dewsb}) is to construct a supersymmetric model admitting a weakly coupled description at long distances, that breaks $SU(2) \times U(1)$ and supersymmetry dynamically. 

Our main example for the short distance theory is SQCD with gauge group $SU(N_c)$ and $N_f$ fundamental flavors $(Q, \t Q)$ in the free magnetic range $N_c+1\leq N_f<\frac{3}{2} N_c$, the ``electric theory''. The low energy theory admits a dual ``magnetic'' description in terms of weakly coupled mesons and baryons, and has a large unbroken symmetry group. After weakly gauging a subgroup and identifying it with the SM gauge group, one of the Higgs fields will arise from the composite meson,
$$
H \subset (Q \t Q)\,.
$$
The magnetic description implies that the dimension of $H$ approaches 1 in the IR. Furthermore, chiral symmetries and supersymmetry forbid explicit mass terms. The role of the technifermion bilinear is now played by a scalar (superfield) bilinear.

More precisely, we postulate that the up-type Higgs $H_u$ is an elementary field, while $H_d$ arises as a composite meson in the SQCD theory --in short, a ``hybrid Higgs'' sector. Other components of this meson field break supersymmetry as in~\cite{Intriligator:2006dd}. The magnetic theory is described by an O'Raifeartaigh model (with composite messengers and direct mediation) plus interactions between the Higgs fields and dimension two messenger operators $\mc O$. The superpotential has the form
\begin{equation}\label{eq:introW}
W_{mag} = W_{O'R} + \lambda_d H_d \mc O_d + \lambda_u H_u \mc O_u\,,
\end{equation}
where $W_{O'R}$ describes the O'Raifeartaigh fields. In our proposal,
the coupling $H_d \mc O_d$ is dictated by Seiberg duality, with $\lambda_d \sim 1$. On the other hand, the interactions of the elementary $H_u$ with the supersymmetry breaking fields are generated by deforming the microscopic theory with an operator $H_u \mc O_u$ that is irrelevant in the UV. While the dimension of $H_u$ is always close to one, at long distance $\mc O_u$ will flow to a dimension 2 operator. This generates a marginal coupling with a naturally small $\lambda_u \ll 1$. 

We will argue that this low energy description of SQCD has the correct structure to simultaneously break supersymmetry and the electroweak symmetry. The dynamical breaking is produced, in the magnetic description, by the Coleman-Weinberg mechanism~\cite{Coleman:1973jx}.

The hypothesis that $H_u$ is elementary and $H_d$ is composite provides a simple explanation for the following points:
\begin{itemize}
\item The hierarchy between the top and bottom/tau masses is generated naturally if $H_d$ is composite (as well as part of the SM matter), while $H_u$ is still taken to be elementary.
\item $H_d$ will have parametrically large mixings with the supersymmetry breaking sector that lead to $m_{H_d}^2 \gg m_{H_u}^2$. As observed in~\cite{Dine:1997qj, Csaki:2008sr}, this helps to solve the $\mu/B_\mu$ problem.
\end{itemize}
From a bottom-up perspective, this possibility appeared e.g. in the more minimal scenarios of~\cite{Cohen:1996vb}, and its relevance for the $\mu/B_\mu$ problem was recently understood in~\cite{Csaki:2008sr}.

Recall that minimizing the tree level Higgs potential requires
\begin{equation}
\begin{aligned}\label{HiggsVac}
\frac{1}{2}m_Z^2 &=  - |\mu|^2 - \frac{m_{H_{u}}^2 \tan^2\beta - m_{H_d}^2 }{\tan^2\beta -1} \cr
\sin 2 \beta  &= \frac{2B_\mu}{2 |\mu^2| + m_{H_u}^2 + m_{H_d}^2} \,.
\end{aligned}
\end{equation}
Generically in gauge-mediated models, the tree-level $\mu$ and $B_\mu$  are forbidden, e.g. by a PQ symmetry, and are dynamically generated at one-loop, implying that $B_\mu \sim 16 \pi^2 \mu^2$. This entails however, that there is no natural solution to the above relations (\ref{HiggsVac}).

The mass hierarchy generated in the  magnetic theory (\ref{eq:introW}) on the other hand is
\begin{equation}\label{eq:intro-hierarchy}
m_{H_u}^2 \ll B_\mu \ll m_{H_d}^2
\end{equation}
and, at the origin of field space, $m_{H_u}^2 < 0$, triggers EWSB. Ref.~\cite{Csaki:2008sr} argued that for adequate scalings of the soft parameters, the hierarchy (\ref{eq:intro-hierarchy}) provides a viable solution to (\ref{HiggsVac}). The theory has an approximate R-symmetry under which the O'Raifeartaigh (tree level) flat direction and $H_d$ have R-charge 2, while $H_u$ has charge -2. This forbids a supersymmetric $\mu$-term and Majorana gaugino masses. The R-symmetry will be broken using the mechanism of~\cite{Essig:2008kz}, producing realistic gaugino and higgsino masses.

Our proposal realizes some of the ideas of~\cite{Cohen:1996vb,Dine:1997qj} and especially~\cite{Csaki:2008sr} in a rather generic SQCD setup.

\subsection{Overview of the model}\label{subsec:overview}

Generating dynamically the electroweak scale does not in principle explain the pattern of masses and mixings of the SM fermions. We adopt the point of view that the same mechanism responsible for breaking $SU(2) \times U(1)$ should also produce the correct flavor textures. Starting from \S \ref{sec:ssector}, the connection between flavor physics and the EW scale is established by combining the hybrid Higgs theory with the single sector models of~\cite{Craig:2009hf}.

We add a field $U$ transforming in the adjoint of the electric gauge group, with a renormalizable $W \sim U^3$ superpotential. The theory confines at a scale $\Lambda$ and generates two types of mesons, $(Q \t Q)$ and $(QU \t Q)$. The first SM generation is identified with components of the dimension 3 meson, the second generation and $H_d$ arise from the dimension 2 meson, while $H_u$ and the third generation fields are elementary. Yukawa couplings are generated at a scale $M_{\rm{flavor}}> \Lambda$ from superpotential couplings between the mesons and Higgs fields. Such operators are irrelevant before the theory confines and become marginal in the infrared. Therefore, different Yukawa couplings are proportional to different powers of the small ratio 
$$
\epsilon=\Lambda/M_{\rm{flavor}}\,,
$$ 
and the correct Yukawa textures are generated. The down-type Yukawas are produced at a higher order in $\epsilon$, explaining the smallness of $m_{bottom}/m_{top}$ dynamically.

In models where unification is possible, the dynamical scale $\Lambda$ is approximately $M_{GUT}$; otherwise it could be smaller. In the class of models analyzed here, achieving unification is difficult. It would be interesting to consider scenarios where this is naturally realized. The scale $M_{\rm{flavor}}$ is roughly an order of magnitude above $\Lambda$, so that $\epsilon \sim 0.1$ generates the correct Yukawa couplings. 

Supersymmetry is now broken by one linear combination of the two mesons, denoted by $X$, which acquires a linear term in the superpotential. From the IR point of view this is again the ISS mechanism~\cite{Intriligator:2006dd}. The model produces composite messengers; composite SM fields couple strongly to the supersymmetry breaking sector and acquire (large) one loop masses. On the other hand, elementary fields acquire their masses predominantly from two loop (direct) gauge mediation. In summary,
\begin{itemize}
\item The first and second generation sfermions and $H_d$ have masses 
\begin{equation}
m_{\t Q_i}^2 \sim m_{H_d}^2  \sim \frac{1}{16 \pi^2} |F_X| \,,
\end{equation}
where $F_X$ is the F-term of the tree level flat direction $X$, and the messenger masses are of order $\sqrt{|F_X|}$.
\item Third generation sfermions and $H_u$ have masses generated from standard gauge mediation at two loops, 
\begin{equation}
m_{GM}^2  \sim \left(\frac{g^2}{16 \pi^2}\right)^2 |F_X| \,,
\end{equation}
\item Gauginos and higgsinos have one loop masses proportional to the VEV of $X$ that breaks the R-symmetry spontaneously, 
\begin{equation}
m_{\psi}  \sim \frac{1}{16 \pi^2} \,|X| \,.
\end{equation}
\end{itemize}

\begin{figure}[t]
\begin{center}
\includegraphics[width=.85\textwidth]{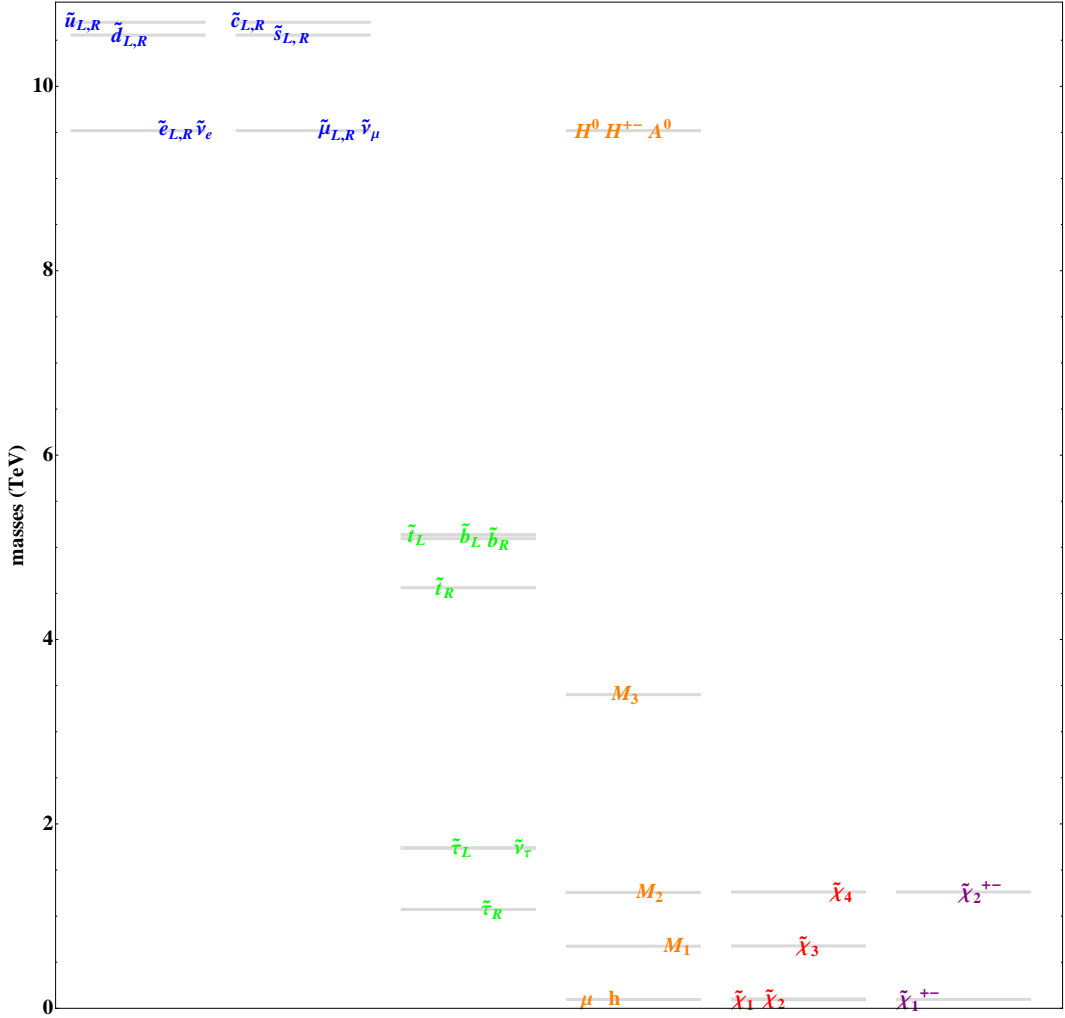}
\caption{Typical spectrum of the model, for $\tan\, \beta \sim 6.5$ and higgsino NLSP. }
\label{fig:Spectrum1}
\end{center}
\end{figure}

Having gaugino and sfermion masses around the TeV sets the supersymmetry breaking scale
$$
\sqrt{|F_X|} \sim\;100 - 200\,{\rm TeV}\,.
$$
This gives a light gravitino
\begin{equation}
m_{3/2} \sim \frac{F_X}{\sqrt{3}M_{\rm Pl}} \sim \mc{O}({\textrm{1--10 eV}})\,.
\end{equation}
A typical spectrum is shown in Figure \ref{fig:Spectrum1}.\footnote{Spectra with inverted hierarchies arise generically in single sector models and they help solve the flavor problem; see~\cite{Craig:2009hf}. Similar phenomenology can arise from a strongly coupled approximately conformal hidden sector, as was recently analyzed in~\cite{Aharony:2010ch}.}

In \S \ref{sec:eft} and \S \ref{sec:pheno} we describe the RG evolution from the messenger scale down to the TeV scale. Models with inverted hierarchies and a hybrid Higgs sector have quite distinct properties, that are studied using a combination of effective potential methods and MSSM RGEs. This will allow us to explicitly obtain the weak scale in terms of the microscopic parameters. \S \ref{sec:pheno} presents more detailed spectra and parameter ranges, and various explicit computations are shown in the Appendix.

\section{Composite Higgs from Seiberg duality}\label{sec:compositeH}

In this section and in \S \ref{sec:dewsb} we present a composite Higgs model that breaks supersymmetry and electroweak symmetry dynamically and calculably. The confining dynamics comes from $SU(N_c)$ SQCD with fundamental flavors, in the free magnetic range. The theory will be analyzed using its weakly coupled description, which we review in \S \ref{subsec:e-m}. The aim is to explain in a simple setup the dynamical mechanism relating supersymmetry breaking to the Fermi scale. It will be argued that this naturally solves the $\mu/B_\mu$ problem, along the lines of~\cite{Dine:1997qj,Csaki:2008sr}. 

As summarized above, in \S \ref{sec:ssector} an extra field transforming in the adjoint of the electric gauge group will be added, in order to generate the Standard Model flavor hierarchies. This modifies the UV properties of the theory, but the long distance description of the electroweak and supersymmetry breaking sectors will be the same as the one for the simpler model analyzed here.

\subsection{Electric and magnetic descriptions}\label{subsec:e-m}

Before introducing the Higgs sector, we start with a brief review of~\cite{Seiberg:1994pq}. The microscopic theory is SQCD with gauge group $SU(N_c)$ and $N_f$ flavors $(Q_i, \tilde Q_j)$ in the free magnetic range $N_c+1 \le N_f < \frac{3}{2} N_c$, with masses $m_i$,
\begin{equation}\label{eq:Wel1}
W_{el}= \sum_{i=1}^{N_f}\, m_i\,Q_i \t Q_i\,.
\end{equation}
The quark masses are ordered $|m_i| >|m_j|$ for $i<j$, and are chosen to be much smaller than the dynamical scale $\Lambda$. This may be generated dynamically, as in~\cite{Dine:2006gm, Essig:2007xk}.

As will be explained shortly, these different quark masses will be required to ensure the stability of our vacuum, which will differ from the ISS model by the addition of extra singlets. Furthermore, at least 5 mass eigenvalues have to be equal, so that there is an unbroken $SU(5)$ global symmetry that can be identified with the SM gauge group. For our purposes, it will be sufficient to have two different eigenvalues,
\begin{equation}
m = \left(\begin{matrix}
m_1 \,\mathbf 1_{N_f-N_c} & 0 \\
0 & m_2\,\mathbf 1_{N_c}
\end{matrix}\right)
\end{equation}
with $|m_1|> |m_2|$. The non-anomalous global symmetries are then
\begin{equation}
SU(N_f-N_c)_V \times SU(N_c)_V \times U(1)_V
\end{equation}
where $Q_i$ ($\t Q_i$) have charge $+1$ ($-1$) under the baryon numer $U(1)_V$.

We weakly gauge a subgroup of the global symmetry group and identify it with the Standard Model gauge group,
\begin{equation}\label{eq:weakgauge1}
SU(3)_C \times SU(2)_L \times U(1)_Y \subset SU(N_c)_V\,.
\end{equation}
We will find it convenient to use an $SU(5)$ ``GUT notation'' as a shorthand for the SM quantum numbers, but no assumption of unification is made.\footnote{Nevertheless, finding single sector models that can also accommodate unification is important. See~\cite{Behbahani:2010wh} for an analysis of this point.} Furthermore, in the realistic models of \S \ref{sec:ssector}, baryon number $U(1)_V$ is automatically gauged; this has the advantage of removing a Nambu-Goldstone boson from the low energy theory.

Below the scale $\Lambda$, the theory has a dual magnetic description~\cite{Seiberg:1994pq} in terms of an SQCD theory with
gauge group
$SU(\tilde N_c \equiv N_f- N_c)$,
singlet mesons $\Phi_{ij}$, and $N_f$ fundamental
flavors $(q_i, \tilde q_j)$. The theory is weakly coupled in the infrared and has superpotential
\begin{equation}\label{eq:Wmag1}
W_{mag} =
h \,{\rm tr}(q \Phi \tilde q)-h \,{\rm tr}(\hat \mu^2\Phi)
\,.
\end{equation}
The magnetic and electric variables are related by
\begin{equation}\label{eq:rel-em}
Q \t Q \sim h \Lambda\, \Phi\;,\;\hat \mu^2 =\left(\begin{matrix}
\mu_1^2 \,\mathbf 1_{N_f-N_c} & 0 \\
0 & \mu_2^2\,\mathbf 1_{N_c}
\end{matrix}\right)\sim \Lambda\,\left(\begin{matrix}
m_1 \,\mathbf 1_{N_f-N_c} & 0 \\
0 & m_2\,\mathbf 1_{N_c}
\end{matrix}\right)\,.
\end{equation}
We choose the superpotential parameters to be real.

The magnetic superpotential of Eq.~(\ref{eq:Wmag1}) (as well as the one in \S \ref{sec:ssector}) receives nonperturbative corrections which, in the regime $|\hat \mu_i| \ll |\Lambda|$, do not affect our results. The model has a classical R-symmetry under which $R(\Phi)=2$, $R(q)=R(\t q)=0$. This symmetry,
which becomes anomalous at the quantum level, will play an important role in what follows. Recall also that $U(1)_V$ is gauged to remove a NG boson.

\subsection{Dynamical supersymmetry breaking}\label{subsec:dsb}

Intriligator, Seiberg and Shih (ISS)~\cite{Intriligator:2006dd} have argued that SQCD, in the free magnetic phase, with small quark masses flows to a weakly coupled O' Raifeartaigh model, thus giving a calculable model with dynamical supersymmetry breaking.

This can be seen from the classical magnetic superpotential Eq.~(\ref{eq:Wmag1}), which breaks supersymmetry by the rank condition, giving nonzero F-terms
\begin{equation}\label{eq:Fij}
\frac{\partial W_{mag}}{\partial \Phi_{ij}}=- h \hat \mu^2_{ij} + h q_i \t q_j \,.
\end{equation}
Parametrizing the fields as
\begin{equation}\label{eq:paramPhi}
\Phi= \left(\begin{matrix} Y_{\tilde N_c \times \tilde N_c} & Z^T_{\tilde N_c \times N_c} \\ \tilde Z_{N_c \times \tilde N_c} &X_{N_c \times N_c}\end{matrix} \right)
\end{equation}
\begin{equation}\label{eq:paramq}
q^T=\left( \begin{matrix} \chi_{\tilde N_c \times \tilde N_c} \\ \rho_{N_c \times \tilde N_c} \end{matrix}\right)\;,\;\tilde q=\left( \begin{matrix} \tilde \chi_{\tilde N_c \times \tilde N_c} \\ \tilde \rho_{N_c \times \tilde N_c} \end{matrix}\right)\,,
\end{equation}
(where $\t N_c=N_f-N_c$ is the rank of the magnetic gauge group) $X$ is flat at tree level (``pseudo-modulus''), while 
\begin{equation}\label{eq:rank-cond}
\langle \chi \t \chi \rangle = \mu^2_1 I_{\t N_c \times \t N_c}\;\;,\;\;V_0=(N_f - \t N_c) (h\mu_2^2)^2\,.
\end{equation}
The minimum corresponds to aligning the nonzero expectation value in the direction $\mu_1^2$ of the largest linear term, while the supersymmetry breaking scale is set by the smaller $\mu_2^2 < \mu_1^2$.

The expectation value for $\chi \t \chi$ completely breaks the magnetic gauge group, leaving
\begin{equation}\label{eq:symm}
SU(\tilde N_c)_G \times SU(\t N_c) \times SU(N_c) \times U(1)_V \to SU(\tilde N_c)_D \times SU(N_c) \times U(1)'\,.
\end{equation}
The fields from $(Y,\chi)$ are supersymmetric at tree level and will not be important in what follows. On the other hand, the $(\rho, Z, X)$ sector gives $N_c$ decoupled O' Raifeartaigh models~\cite{Intriligator:2006dd}. In particular, the $(\rho, Z)$ fields couple to the tree level F-terms and have supersymmetric masses of order $h \mu_1$ and nonsupersymmetric splittings of order $h \mu_2$. Once $g_{SM} \neq 0$ in (\ref{eq:weakgauge1}), they play the role of composite messengers, generated dynamically by the theory. 

As will be reviewed in \S \ref{sec:dewsb}, the flat directions are stabilized at $\langle X \rangle =0$ once one loop effects from $\rho$ and $Z$ are taken into account. Nonperturbative effects, which have been neglected here, create supersymmetric vacua. As long as $\mu_i \ll |\Lambda|$, the metastable vacuum is parametrically long lived. Notice that the $U(1)_R$ symmetry is unbroken; in \S \ref{subsec:mu} the superpotential will be deformed and the R-symmetry will be broken explicitly and spontaneously~\cite{Essig:2008kz}.

\subsection{A hybrid Higgs sector}\label{subsec:hybrid}

In order to generate the small ratio $M_Z/M_{Pl}$ dynamically, we postulate that some of the Higgs fields arise as composites of the SQCD theory introduced before. We take $H_u$ to be an elementary field, while $H_d$ is generated from the meson $Q \t Q$. In the weakly coupled magnetic description the dynamical breaking of $SU(2) \times U(1)$ will be fully calculable, arising as a tree level plus one loop effect.

After weakly gauging the SM gauge group, the mesons $Q \t Q$ contain ${\bf 5}+ \overline {\bf 5}$ representations, which have the correct quantum numbers for a Higgs field. The global symmetry is broken to (\ref{eq:symm}) and $SU(5)_{SM}$ is embedded into the unbroken subgroup $SU(N_c)$~\cite{Essig:2008kz}. This means that $H_d$ is identified with a $\overline {\bf 5}$ component from $X$,
\begin{equation}
H_d \equiv X_{\overline {\bf 5}}\,.
\end{equation}
It is important that this element comes from an off-diagonal component of the meson that does not have a tree level F-term. 

Anomaly cancellation requires adding an elementary ``spectator'' $S_{\overline {\bf 5}}$ with the quantum numbers of $H_d$. Notice that $X$ also contains conjugate representations that could couple to $H_d$ once more general electric perturbations are turned on (see \S \ref{sec:dewsb}). This is avoided by coupling such extra unwanted matter to the spectator,
\begin{equation}\label{eq:DelWel}
\Delta W_{el}= \lambda S_{\overline {\bf 5}}  (Q \t Q)_{{\bf 5}}\,,
\end{equation}
as in~\cite{Franco:2009wf}. In the magnetic theory, this gives masses $\lambda \Lambda$ to the unwanted representations. Such deformations can affect supersymmetry breaking in important and interesting ways, as we now discuss (see e.g.~\cite{Green:2010ww,Behbahani:2010wh} for recent analysis).

Below the high energy scale $\lambda \Lambda$, $S_{\overline {\bf 5}}$ and the meson component $(Q \t Q)_{{\bf 5}}$ can be integrated out. In the undeformed theory the F-term (\ref{eq:Fij}) for $(Q \t Q)_{{\bf 5}}$ required $\rho_{\overline {\bf 5}} \t \rho_{\bf 1} = 0$. However, due to the deformation (\ref{eq:DelWel}), this field is no longer part of the low energy spectrum and we have no such constraint.~\footnote{Equivalently, if the heavy fields are not integrated out, it is possible to turn on $\rho_{\overline {\bf 5}} \t \rho_{\bf 1}$ along an F-flat direction if $\Lambda S_{\overline {\bf 5}}=\rho_{\overline {\bf 5}} \t \rho_{\bf 1}$.} Let us analyze the modified F-term conditions. Besides the vacuum configuration presented in \S \ref{subsec:dsb}, there are now new ways of cancelling the linear F-terms by turning on $\rho_{\overline {\bf 5}}$ and $\t \rho_{\bf 1}$. Taking into account the rank condition, consider
\begin{equation}\label{eq:new}
\chi \t \chi  = \mu_1^2 \mathbf 1_{\t N_c-1}\;\;,\;\; \rho_{\bf 1} \t \rho_{\bf 1} =\rho_{\overline {\bf 5}} \t \rho_{\bf 5}= \mu_2^2\,.
\end{equation}
Writing
\begin{equation}
|\rho_{\bf 1}|= \eta\,\mu_2\;\;,\;\;|\t \rho_{\bf 1}|= \eta^{-1} \mu_2\;\;,\;\;
|\rho_{\overline {\bf 5}}|= \xi^{-1}\,\mu_2\;\;,\;\;|\t \rho_{\bf 5}|= \xi \mu_2\,,
\end{equation}
the F-term potential now becomes
\begin{equation}\label{eq:VF}
V_F = (h \mu_1^2)^2+ (N_f - \t N_c -2) (h \mu_2^2)^2+ \eta^2\xi^2 (h \mu_2^2)^2\,.
\end{equation}
Here, two F-terms from $X$ have been cancelled by turning on the $\rho_i$, at the price of a new uncancelled F-term in the $Y$ direction. The last term in the potential comes from the off-diagonal $W_{\Phi_{\bf 5}}= h \rho_{\bf 1} \t \rho_{\bf 5}$.\footnote{We thank N. Craig for useful discussions on these points.}

First consider the global limit, where $SU(5)_{SM} \times U(1)_V$ are not gauged. Then the potential (\ref{eq:VF}) has a runaway $\eta \to 0$ and $\xi \to 0$. Of course, once some of the expectation values become comparable to the dynamical scale, the above analysis breaks down and microscopic effects need to be included. If we ignore these corrections, the runaway direction has an F-term energy
\begin{equation}
V_{runaway}=(h \mu_1^2)^2+ (N_f - \t N_c -2) (h \mu_2^2)^2\,.\label{eq:energychiralvac}
\end{equation}
The ISS metastable vacuum has lower energy as long as
\begin{equation}\label{eq:cond1}
\mu_2^2 \le \frac{1}{\sqrt{2}} \mu_1^2\,.
\end{equation}
Thus, already order one differences in the quark masses stabilize the hybrid Higgs vacuum against decays towards the new runaway created by the deformation (\ref{eq:DelWel}).

Having understood the global limit, let us take into account the effects of weakly gauging $SU(5)_{SM} \times U(1)_V$. This gives a  D-term potential
\begin{equation}\label{eq:VD}
V_D = \frac{1}{2} \mu_2^4\left( g_{SM}^2 \left(\xi^2-\frac{1}{\xi^2}\right)^2+ g_V^2 \left(\xi^2-\frac{1}{\xi^2}+\eta^2-\frac{1}{\eta^2}\right)^2\right)\,.
\end{equation}
Combining (\ref{eq:VF}) and (\ref{eq:VD}), $\xi$ and $\eta$ are stabilized away from the origin. For realistic values of the couplings $g^2 \sim \mc O(1)$, the minima are at $(\xi, \eta) \lesssim \mc O(1)$. The condition (\ref{eq:cond1}) is then relaxed, such that even percent-level differences between $\mu_1^2$ and $\mu_2^2$ suffice to make the vacuum energy of the new configuration larger than the ISS value.

It should be noted that, while different masses $\mu_i^2$ make the hybrid Higgs construction fully stable against decay towards the above new vacua, in applications to single-sector models as in \S\ref{sec:ssector} it will be useful to consider degenerate masses in order to obtain realistic spectra. In this case, the decay channel to (\ref{eq:new}) is allowed, and one has to ensure that the ISS vacuum is long-lived enough. As will be explained in \S\ref{subsec:newvacua}, in these single-sector models there are new vacua close to the origin in field-space, in which vector spectators acquire VEVs, and which have lower energy than those involving chiral spectators. The decay rate of the ISS vacuum is expected to be dominated by the tunneling towards the lower lying minima, and an explicit calculation of the bounce action corresponding to the semiclassical tunneling configurations shows that as long as $h \lesssim \mc O(1)$, the metastable vacuum is numerically (though not parametrically) long-lived. We will come back to this point in \S\ref{subsec:newvacua}.

In summary, we will choose $\mu_1^2 \gtrsim \mu_2^2$ so that the state (\ref{eq:new}) is energetically disfavored; in this case we may safely turn on the deformation (\ref{eq:DelWel}) and work in the metastable vacuum described in \S \ref{subsec:dsb}. Denoting
\begin{equation}\label{eq:defXp}
X= X' \oplus H_d \oplus X_{\bf 5}
\end{equation}
(where $X'$ has an even number of ${\bf 5}+ \overline {\bf 5}$), the low energy theory will contain only $X'$ and $H_d$. The interactions between $H_d$ and the supersymmetry breaking sector are fully determined by Seiberg duality, and follow from the superpotential (\ref{eq:Wmag1}).

The final step in defining the Higgs sector is to specify the interactions between the elementary $H_u$ and the SQCD fields. We postulate that at some high scale $M_H > \Lambda$ there is new physics that generates the irrelevant interaction
\begin{equation}\label{eq:WH}
W_H \sim \frac{1}{M_H^2} (Q \t Q) H_u (Q \t Q)\,.
\end{equation}
In fact, we will see in \S \ref{sec:ssector} that these interactions are also required for producing the SM flavor structure once some of the SM fermions are generated through compositeness. Then $M_H$ will be identified with the ``flavor scale'' $M_{{\rm flavor}}$, at which the SM Yukawa couplings are produced.

In the low energy theory, (\ref{eq:WH}) gives cubic terms of the form $W_{mag} \sim X H_u X + Z H_u \t Z$. The trilinear couplings involving the traceless part of $X$ (which are responsible for the SM Yukawas of composite generations) will be analyzed in \S \ref{sec:ssector}. Their effect on EW physics starts at two loops. In contrast, in order to preserve the supersymmetry breaking structure of \S \ref{subsec:dsb}, no extra couplings to the singlet $\tr\,X$ should be generated. This can be enforced at the scale $M_H$ by an approximate discrete symmetry under which $\tr X$ is odd while the rest of the SM fields are even. We focus then on the terms $Z H_u \t Z$ that couple $H_u$ to the composite messengers. The contribution to the magnetic superpotential is
\begin{equation}
W_{mag} \supset \lambda_u \tr(Z H_u \t Z)\;\;,\;\;\lambda_u \sim \mc O \left( \frac{\Lambda^2}{M_H^2}\right)\,.
\end{equation}
Since $M_H > \Lambda$, this naturally gives $\lambda_u \ll 1$.

To summarize, the magnetic description of the supersymmetry breaking plus Higgs sectors becomes
\begin{equation}\label{eq:Wmag2}
W_{mag}= \left[-h \mu_2^2\,\tr X' + h\,\tr(\rho X' \t \rho) + h \mu_1 \tr (\rho \t Z + \t \rho Z)\right]+ \lambda_d \tr(\rho H_d \t \rho)+ \lambda_u \tr(Z H_u \t Z)\,.
\end{equation}
For clarity, the supersymmetry breaking fields have been grouped inside the square brackets. We have already expanded in the fluctuations (\ref{eq:paramPhi}), (\ref{eq:paramq}), and dropped inconsequential contributions from the supersymmetric fields $(Y, \chi)$. Recall that $X'$ was defined in Eq.~(\ref{eq:defXp}); when there is no confusion we will drop the primes in this field.

The $H_d$ interaction is generated by $h \tr(q \Phi \t q)$ in (\ref{eq:Wmag1}), and $\lambda_d=h$. On the other hand, the interaction $Z H_u \t Z$ with the elementary Higgs follows from the dimension 5 operator (\ref{eq:WH}) added to the electric theory. Therefore
\begin{equation}\label{eq:luld}
\lambda_u \ll \lambda_d\,
\end{equation}
and there is a parametrically strong mixing between the composite Higgs $H_d$ and the supersymmetry breaking sector, while the elementary field couples through a highly suppressed operator (and decouples in the limit $M_H/\Lambda \to \infty$).

The R-symmetry assignments given at the end of \S \ref{subsec:e-m}, together with the superpotential (\ref{eq:Wmag2}), imply that the Higgs fields have R-charges
\begin{equation}\label{eq:R-assign}
R(H_u)=-2\;,\;R(H_d)=2\,.
\end{equation}
This symmetry forbids a supersymmetric $\mu$ term $W=\mu H_u H_d$.

\section{Dynamical electroweak symmetry breaking}\label{sec:dewsb}

We are now ready to analyze the breaking of $SU(2) \times U(1)$ in the magnetic theory, and establish its relation to supersymmetry breaking.

First, at tree level the theory with superpotential (\ref{eq:Wmag2}) exhibits supersymmetry breaking as in the ISS model described in \S \ref{subsec:dsb}. This happens because we have not added interactions involving the supersymmetry breaking fields corresponding to the diagonal components of $X$. Both $H_u$ and $H_d$ are flat at the classical level.

To understand the dynamics of the low energy modes, one loop effects from the heavy fields $(\rho, Z)$ have to be taken into account. Keeping $X$, $H_u$ and $H_d$ as background fields and integrating out the messengers produces a Coleman-Weinberg potential~\cite{Coleman:1973jx},
\begin{equation}\label{eq:CW}
V_{CW}=\frac{1}{64 \pi^2}\,{\rm Str} \;\mc M^4\,\left(\log \frac{\mc M^2}{(h  \mu_1)^2}-\frac{3}{2}\right)\,.
\end{equation}
For computational purposes, it is simpler to evaluate the potential at the messenger scale $h \mu_1$. However, the dependence on the scale cancels out, and the one loop potential actually gives finite effects independent of $h \mu_1$.

The fermion mass matrix can be read off from (\ref{eq:Wmag2}),
\begin{equation}\label{eq:Mf}
W_{mag} \supset {\rm tr}\;\left( \begin{matrix} \rho& Z\end{matrix} \right) \left( \begin{matrix}  h X + \lambda_d H_d & h \mu_1 \\ h \mu_1 &\lambda_u H_u\end{matrix} \right) \left( \begin{matrix} \tilde \rho\\ \tilde Z\end{matrix} \right)\;\;,\;\;M_f=\left( \begin{matrix}   h X + \lambda_d H_d & h \mu_1 \\ h \mu_1 & \lambda_u H_u\end{matrix} \right)\,.
\end{equation}
The bosonic mass matrix includes off-diagonal F-terms $F_X^*= h \mu_2^2$ as well. Details of the calculation (\ref{eq:CW}) can be found in Appendix \ref{app:one-loop}.

\subsection{One-loop effects and electroweak symmetry breaking}\label{subsec:EW}

Expanding the CW potential to quadratic order in the fluctuations $(X, H_u, H_d)$,
\begin{equation}\label{eq:CWH}
V_{CW}=m_X^2 |X|^2+m_{H_u}^2 |H_u|^2 + m_{H_d}^2 |H_d|^2-B_\mu (H_u H_d + c.c.)+  \ldots
\end{equation}
gives
\begin{eqnarray}\label{eq:softmasses}
m_X^2&=&\frac{h^2\t N_c}{8 \pi^2}(\log 4 -1)\, \frac{h^2 \mu_2^4}{\mu_1^2}\,,\nonumber\\
m_{H_d}^2&=&\frac{\lambda_d^2\t N_c}{8 \pi^2}(\log 4 -1)\,\frac{h^2 \mu_2^4}{\mu_1^2}\,, \nonumber\\
m_{H_u}^2&=&-\frac{\lambda_u^2\t N_c}{8 \pi^2}(1-\log 2)\,\frac{h^2 \mu_2^4}{\mu_1^2}\,,\nonumber\\
B_{\mu}&=&\frac{\lambda_u \lambda_d\t N_c}{8 \pi^2}(1-\log 2)\,\frac{h^2 \mu_2^4}{\mu_1^2}\,.
\end{eqnarray}
The mass contribution $m_X^2$ was found in~\cite{Intriligator:2006dd} and stabilizes the pseudo-modulus $X$ at the origin. The squared Higgs masses are generated one loop factor below the supersymmetry breaking scale, and are suppressed by the cubic couplings $\lambda_u$ and $\lambda_d$. This gives a hierarchy
$$
m_{H_u}^2 \ll B_\mu \ll m_{H_d}^2\,.
$$
Here we are neglecting two loop gauge mediated effects, that will be included in the full model of \S \ref{sec:eft}. Note that for real values of the high energy parameters, no phases appear in the soft masses given above.

Importantly, integrating out the heavy messengers produces a tachyonic mass for $H_u$, $m_{H_u}^2<0$, so that electroweak symmetry is spontaneously broken. The up-type Higgs couples to the meson messengers $(Z, \t Z)$, as opposed to the pseudo-modulus $X$ that is stabilized at the origin through its coupling to  $(\rho, \t \rho)$. The destabilization of the origin $H=0$ is a nonperturbative phenomenon from the point of view of the original electric theory, and appears in the magnetic description as a one loop effect produced by the composite messengers. We conclude that the effective model (\ref{eq:Wmag2}) that appears as the long distance description of our SQCD theory has the right structure to break the electroweak symmetry (and supersymmetry) dynamically.

This provides an alternative mechanism to radiative EWSB~\cite{REWSB}, where $m_{H_u}^2$ becomes negative due to the RG evolution driven by the top quark Yukawa coupling. Radiative EWSB is particularly important at small $\tan \beta$ and when supersymmetry is broken at a high scale. Our mechanism could become useful outside those regimes. In fact, in the realistic models presented below, we will break $SU(2) \times U(1)$ by a combination of dynamical and radiative effects.

The position of the EWSB vacuum is obtained by minimizing (\ref{eq:CWH}) plus the quartic potential coming from the SM gauge coupling D-terms,
\begin{equation}
V_D= \frac{g^2}{2}|H_u^+ H_d^{0*}+H_u^0 H_d^{-*}|^2+\frac{1}{8}(g^2+g'^2)\left(|H_u^0|^2-|H_d^0|^2+|H_u^+|^2-|H_d^-|^2 \right)^2\,.
\end{equation}
The CW potential (\ref{eq:CW}) also generates quartic Higgs couplings of order $\lambda_{u,d}^4/(32 \pi^2)$, whose effect is negligible compared to the D-term contribution. 

Given the (real valued) soft parameters of Eq.~(\ref{eq:softmasses}), there are vacuum solutions with $\langle H_d^- \rangle =\langle H_u^+\rangle=0$ and with real values for $\langle H_u^0 \rangle$ and $\langle H_d^0 \rangle$. First, extremizing with respect to $H_d$ gives
\begin{equation}\label{eq:tanbeta1}
\tan \beta \approx \frac{m_{H_d}^2}{B_\mu} \sim \frac{\lambda_d}{\lambda_u}\,.
\end{equation}
Then the $H_u$ extremum gives, in terms of $v^2=H_u^2+H_d^2$,
\begin{equation}\label{eq:v1}
\frac{1}{4}(g^2+g'^2)v^2= \frac{m_{H_d}^2-\tan^2 \beta\,m_{H_u}^2}{\tan^2 \beta -1}\,.
\end{equation}
Using (\ref{eq:softmasses}), all the terms in this equation are of the same order, and give
\begin{equation}\label{eq:v2}
\frac{1}{4}(g^2+g'^2)v^2 \approx \frac{\lambda_u^2}{16 \pi^2} \frac{h^2\t N_c \mu_2^4}{\mu_1^2}\,.
\end{equation}
The dynamically generated electroweak scale is then one loop below the supersymmetry breaking scale and is controlled by the coupling of $H_u$ to the messenger sector. 

The model then gives a decoupling limit of the MSSM, having large $\tan \beta$ and $|m_{H_d}| \gg |m_{H_u}|$. Notice that all the terms contributing to (\ref{eq:v1}) are naturally of the same order. For example, a supersymmetry breaking scale
$$
h \mu_i \sim \mathcal O (100\,{\rm TeV})
$$
and $\lambda_u \sim \mathcal O(0.01)$ give a Fermi scale of the correct magnitude. It is necessary to point out that in the concrete model presented below we will need larger values $\lambda_u \sim 0.1$ in order to get a realistic spectrum. Stop effects will also become important, and some amount of tuning will be required.

\subsection{R-symmetry, $\mu$ term and gaugino masses}\label{subsec:mu}

The calculations of Appendix \ref{app:one-loop} reveal that there is no one loop $\mu$ term from integrating out the messengers; similarly, no Majorana masses for gauginos are generated. The reason is that in the limit $g_{SM} \to 0$ the supersymmetry breaking vacuum preserves the R-symmetry defined before. For $g_{SM} \neq 0$ the Higgs VEVs spontaneously break this symmetry, but such breaking generates negligibly small higgsino and gaugino masses. An extra source of R-symmetry breaking is hence needed.

This is remedied as follows~\cite{Essig:2008kz}. The electric theory is perturbed by a quartic operator produced at some high scale $\Lambda_0$,
\begin{equation}\label{eq:muphiWel}
\Delta W_{el}=\frac{\lambda}{2\Lambda_0} (Q \t Q)^2\;\;,\;\;\Lambda_0 \gg \Lambda\,.
\end{equation}
At long distance this becomes a relevant mass term with a naturally small coefficient,
\begin{equation}\label{eq:muphiW}
\Delta W_{mag}=\frac{1}{2}h^2\muphi\,\tr\, \Phi^2\;\;,\;\;\muphi\equiv \frac{\Lambda^2}{\Lambda_0}\lambda\,.
\end{equation}
This term breaks $U(1)_R$ explicitly because $R(\Phi)=2$. We refer the reader to~\cite{Giveon:2007ef} for an analysis of general polynomial deformations of this theory.

The deformation by $\Phi^2$ creates supersymmetric vacua at a distance $\Phi \sim \hat \mu^2/\muphi$. However, in the limit $\muphi \ll |\hat \mu|$, the metastable vacuum found before still exists, albeit shifted away from the origin~\cite{Essig:2008kz},
\begin{equation}\label{eq:Xvev}
\langle h X \rangle\sim 16 \pi^2\,\muphi \,\frac{\mu_1^2}{\t N_c\mu_2^2}\,.
\end{equation}
The expectation value of $X$ is larger than $\muphi$ by an inverse loop factor because the vacuum arises by balancing the tree level tadpole $\mu_2^2 \muphi X$ against the one loop contribution $m_X^2|X|^2$. In the limit $\muphi/\mu_i \ll 1$ the supersymmetry breaking vacuum can be made parametrically long-lived.

The fact that the spontaneous breaking of the R-symmetry ($\langle h X \rangle$) dominates over the explicit breaking ($\muphi$) implies that large enough gaugino and higgsino masses can be generated even if $\muphi/\mu_i$ is small. Gaugino mass unification occurs as a consequence of the global symmetry respected by the messengers. For instance,~\cite{Essig:2008kz} found that for small $X/ \mu_i$, gaugino masses are
\begin{equation}\label{eq:mgaugino}
m_{\lambda_a}\sim \frac{g_a^2\t N_c}{16 \pi^2} \langle h X \rangle \,\left(\frac{F}{M^2}\right)^3\sim g_a^2 \muphi\,\frac{\mu_2^4}{\mu_1^4}\,,
\end{equation}
where $F= h^2 \mu_2^2$, $M^2= h^2 \mu_1^2$. TeV scale gaugino masses can be obtained for $\muphi \sim \rm{TeV}$. 

Importantly for our purposes, in the presence of (\ref{eq:muphiW}) a nonzero one loop $\mu$ term is produced by the heavy messengers (see Appendix \ref{subsec:appmu}),
\begin{equation}\label{eq:mu}
\mu \sim \frac{\lambda_u \lambda_d\t N_c}{16 \pi^2} \langle h X \rangle \,\left(\frac{F}{M^2}\right)^3\sim \lambda_u \lambda_d \muphi\,\frac{\mu_2^4}{\mu_1^4}\,.
\end{equation}
Our proposal naturally leads to a small higgsino mass, proportional to $\lambda_u$. For instance, for $\muphi$ in the TeV range, $\t N_c=1$, $\lambda_u \sim 0.01$ and $\lambda_d \sim 2\pi$, we obtain $\mu_1 \sim \mu_2 \sim 100\;{\rm GeV}$.\footnote{In the realistic proposals in \S \ref{sec:eft} and \S \ref{sec:pheno}, we will focus on values $\lambda_d \sim 1$ and $\lambda_u \sim 0.1$.} Finally, $\langle h X \rangle \neq 0$ also corrects the soft parameters computed in Eq.~(\ref{eq:softmasses}), in a way which will be discussed in the following sections.

\subsection{Comments on the $\mu/B_\mu$ problem}\label{subsec:muBmu}

In gauge mediation, the $\mu/B_\mu$ problem appears because parameters with different mass dimension are generated at the same loop level. Typically, loop effects from the hidden sector produce $\mu^2 \ll B_\mu$. EWSB is then either impossible (if $m_{H_u} \sim m_{H_d} \sim \mu$), or extremely fine-tuned (if $\mu$ is at the weak scale and $m_{H_u}^2 \sim m_{H_d}^2 \sim B_\mu$). In general, this is addressed using new mechanisms that ensure
$$
m_{soft}^2 \sim \mu^2 \sim B_\mu \sim m_{H_{u,d}}^2\,.
$$
See e.g.~\cite{Dvali:1996cu, Komargodski:2008ax} and references therein.

On the other hand, the composite Higgs model produces soft parameters (\ref{eq:softmasses}) and (\ref{eq:mu}) with
$$
m_{H_u}^2 \ll B_\mu \ll m_{H_d}^2\,.
$$
This gives a viable electroweak scale, with all the terms in Eq.~(\ref{eq:v1}) being of the same order of magnitude. Therefore, the hierarchy $\mu^2 \ll B_\mu$ is no longer problematic, and in fact it leads to an attractive phenomenology that we analyze below. From the low energy theory this follows from the strong mixing between $H_d$ and the ``hidden'' sector. In fact, the microscopic theory shows that $H_d$ is part of the supersymmetry breaking sector, and Seiberg duality provides a weakly coupled dual where such dynamics can be understood.

Finally, it is necessary to point out that the simple mechanism presented so far receives corrections from interactions with the rest of the MSSM fields, especially with the top and stop. Such effects are analyzed in \S \ref{sec:ssector} -- \S \ref{sec:pheno}. Constructing a realistic single sector model places certain constraints on the parameters of the Higgs sector, which lead to some amount of fine-tuning (as discussed in Appendix \ref{subsec:tuning}).

\section{Building a composite supersymmetric SM}\label{sec:ssector}

After having specified the supersymmetry breaking and Higgs sector, we next consider the SM generations and focus on the question of how the Yukawa textures arise. Our goal in this second part of the work is to present fully realistic models and analyze the RG evolution from the messenger scale down to the top mass scale.

We advocate the idea that the dynamics responsible for breaking $SU(2) \times U(1)$ should also generate the flavor hierarchies. This will be accomplished by combining the previous mechanism of EWSB with the single sector model of~\cite{Craig:2009hf}. We should point out that the idea of SM fermions composed of preons that also break the weak symmetry has been explored over the years using different tools; see~\cite{Kaplan:1991dc} and references therein.

In our proposal, generating realistic flavor textures requires adding a field that transforms in the adjoint representation of the gauge group. We start by discussing this new theory and its long distance description. Then in \S \ref{subsec:ssmodel} we review how this theory gives rise to realistic Yukawa couplings via compositeness, and we go on to show that the mechanism for dynamical EWSB can also be applied in this context, with certain modifications that we shall analyze. Finally, \S \ref{subsec:summary}  summarizes the main properties of the model and its soft parameters.

\subsection{The microscopic theory and its magnetic dual}

The electric theory is $SU(N_c)$ SQCD with $N_f$ flavors $(Q, \t Q)$ and a field $U$ in the adjoint representation of the gauge group; this has been studied in~\cite{Kutasov:1995ve, Kutasov:1995np, Kutasov:1995ss}. A general renormalizable superpotential for $U$ is introduced,
\begin{equation}\label{eq:adj-Wel1}
W_{el}= \frac{g_U}{3} \,\Tr U^3 + \frac{m_U}{2} \,\Tr U^2 \,.
\end{equation}
Here `$\Tr$' means a trace over the gauge indices, while `$\tr$' will be reserved, as before, for traces over flavor indices. 

The cubic superpotential restricts the chiral ring mesons to 
\begin{equation}
M_1 \equiv (Q \t Q)\;,\;M_2\equiv (Q U \t Q)
\end{equation}
which will be identified with the first two SM composite generations. We will shortly perturb the superpotential using the operators $Q \t Q$ and $Q U \t Q$ to produce supersymmetry breaking vacua.

Below the dynamical scale $\Lambda$, the theory admits a magnetic description in terms of an $SU(\t N_c \equiv 2 N_f-N_c)$ gauge group, with $N_f$ fundamental quarks $(q, \t q)$, singlets $M_1$ and $M_2$, and an adjoint $\t U$ of the magnetic gauge group. The superpotential includes a cubic polynomial in $\t U$ plus cubic and quartic interactions between the magnetic quarks, the singlet fields and $\t U$.

This theory is IR free and stable in the range 
$$
\frac{N_c}{2}<N_f< \frac{2}{3}N_c\,.
$$
The magnetic theory simplifies for
\begin{equation}\label{eq:Nf}
2N_f-N_c=1\,,
\end{equation}
since there is no magnetic gauge group. For simplicity, unless the formulae include explicit factors of $\t N_c$, in what follows we restrict to the case $\t N_c=1$, although our construction can be applied to the full free magnetic range. The magnetic quarks correspond then to the dressed baryons
\begin{equation}
q = \frac{[Q]^{N_f}[UQ]^{N_f-1}}{\Lambda^{3N_f-2}}\;\;,\;\;\t q = \frac{[\t Q]^{N_f}[U \t Q]^{N_f-1}}{\Lambda^{3N_f-2}}\,.
\end{equation}
It is also convenient to redefine the singlet mesons to have dimension one,
\begin{equation}
\Phi_1 = \frac{(Q \t Q)}{\Lambda}\;\;,\;\;\Phi_2 = \frac{(Q U\t Q)}{\Lambda^2}\,.
\end{equation}

In terms of these fields, the classical magnetic superpotential is
\begin{equation}\label{eq:WmagU}
W_{mag}=h\, \tr\,q \left(-\frac{1}{2}(N_c-1)\frac{m_U}{g_U \Lambda} \Phi_1+ \Phi_2 \right) \t q
\end{equation}
(plus a nonperturbative contribution which is negligible for our analysis).The appearance of the ratio $m_U/\Lambda$ multiplying $\Phi_1$ can be understood in terms of a non-anomalous R-symmetry ($R(U)=2/3$) that is restored in the limit $m_U/\Lambda \to 0$. The long distance theory consists then of magnetic quarks $(q, \t q)$ coupled to the linear combination of mesons appearing in (\ref{eq:WmagU}), plus a free meson corresponding to the orthogonal combination.

\subsection{Supersymmetry breaking}\label{subsec:dsbU}

The next step is to introduce appropriate deformations to generate supersymmetry breaking vacua. Notice that the IR theory reduces to the one discussed in \S \ref{subsec:e-m} after identifying
\begin{equation}\label{eq:Phidef}
 \Phi \equiv -\frac{1}{2}(N_c-1)\frac{m_U}{g_U \Lambda} \Phi_1+ \Phi_2\,.
\end{equation}
Supersymmetry and R-symmetry breaking will ensue once linear and quadratic deformations in $\Phi$ are added to the magnetic superpotential.

There are, however, two important differences with the theory of \S \ref{sec:compositeH}:
\begin{enumerate}
\item There is an extra meson which we denote by $\t \Phi$ (the orthogonal combination to (\ref{eq:Phidef})) that is decoupled from the supersymmetry breaking sector at the classical level. Once supersymmetry is broken, this direction could become tachyonic. We have to make sure that the magnetic deformations induce a large enough mass.
\item The UV theory is different. In particular, the mesons and baryons have different UV dimensions than the ones in \S \ref{sec:compositeH}. This is important for our purposes because the flavor hierarchies will be generated at a scale $M_{{\rm flavor}}>\Lambda$. While supersymmetry breaking is an IR effect, driven by the combination $\Phi$, for the purpose of computing the SM Yukawa couplings we have to keep track of mesons with different dimensions.
\end{enumerate}
In order to analyze the supersymmetry breaking vacuum, here we work in terms of the fields $\Phi$ and $\t \Phi$, but starting from the next subsection we reintroduce $\Phi_1$ and $\Phi_2$ in connection to the SM Yukawa matrices.

Let us deform the electric theory by a polynomial in the mesons,
\begin{equation}\label{eq:DelWU}
\Delta W_{el}= m (Q \t Q)+ \lambda_1 (QU \t Q)+ \frac{\lambda_2}{2 \Lambda_0} (Q \t Q)^2+ \ldots
\end{equation}
We require that $m \ll \Lambda$ and $\lambda_1 \ll 1$, and the quartic coupling is generated at a scale $\Lambda_0 \gg \Lambda$.
The magnetic superpotential becomes
\begin{equation}\label{eq:WmagU2}
W_{mag} =\left[ -h \,\tr (\hat \mu^2\Phi)+ h \tr (q \Phi \t q) + \frac{1}{2}h^2 \tr(\muphi\Phi^2) \right] -h \,\tr (\t \mu^2 \t \Phi) + \frac{1}{2}h^2 \,\tr(\widetilde{ \muphi}\t \Phi^2)+\ldots
\end{equation}
The largest of $m/\Lambda$ and $\lambda_1$ gives the leading contribution to the linear terms in the IR theory; here the relation between electric and magnetic parameters is analogous to (\ref{eq:rel-em}). 

Different electric quark masses or cubic couplings imply that $\hat \mu^2$ is an $N_f \times N_f$ matrix. For our purposes it will be enough to consider the matrix structure given in (\ref{eq:rel-em}). Similarly, the mass terms follow from the quartic and higher order deformations in (\ref{eq:DelWU}). For simplicity, $\muphi$ will be taken to be proportional to the identity matrix. The couplings with and without tildes are of the same order of magnitude --they differ by order one numerical factors that enter into the definition of $(\Phi, \t \Phi)$ in terms of $(\Phi_1, \Phi_2)$. Also, mixed terms $\Phi \t \Phi$ are not included because their effect can be absorbed into a redefinition of $\muphi$ and $\hat \mu$ after stabilizing $\t \Phi$ (see below).

The terms grouped inside the square brackets give the model discussed in the first part of the work. There is a supersymmetry breaking direction $X$ corresponding to the lower sub-block of $\Phi$ (see Eq.~(\ref{eq:paramPhi})), as well as composite messengers $(\rho, Z)$. As in \S \ref{subsec:mu}, R-symmetry is broken both explicitly and spontaneously, with the latter dominating. For what follows, it is important to keep in mind that $X$ and $Z$ arise now from a combination of dimension 2 and dimension 3 fields in the UV theory. The Coleman-Weinberg potential for the fields with definite UV dimension reads
\begin{equation}\label{eq:CWU}
V_{CW}=m_{CW}^2 \left| -\frac{1}{2}(N_c-1)\frac{m_U}{g_U \Lambda} \Phi_1+ \Phi_2 \right|^2+ \ldots\,,
\end{equation}
where the one loop mass $m_{CW}^2=m_X^2$ was defined in Eq.~(\ref{eq:CWH}).

On the other hand, since the meson $\t \Phi$ is not coupled to $(q, \t q)$, the rest of the terms imply that $\t \Phi$ is stabilized supersymmetrically at $\langle h \t \Phi \rangle = \t \mu^2/\widetilde{\muphi}$. We conclude that the same deformations that break supersymmetry also stabilize $\t \Phi$ and resolve the issue of potential tachyons from microscopic corrections. This then gives a consistent model of supersymmetry and R-symmetry breaking in SQCD with fundamentals and an adjoint.\footnote{We refer the reader to~\cite{Ooguri:2006pj, Amariti} for a different construction of metastable vacua in SQCD with an adjoint.}

\subsection{Flavor textures and composite Higgs}\label{subsec:ssmodel}

Finally, we add in the Higgs sector of \S \ref{sec:compositeH} and combine with the composite MSSM of~\cite{Craig:2009hf}. The matter content of the model is the following:
\begin{itemize}
\item The supersymmetry breaking sector is composite and arises from the diagonal components of (\ref{eq:Phidef}) plus magnetic quarks.
\item $H_u$, the third generation ($\Psi_3$), and gauge supermultiplets are elementary.
\item $H_d$ and the second generation arise as dimension 2 composites from $Q \t Q$.
\item The first generation is generated from the dimension 3 meson $QU \t Q$\,.
\end{itemize}

\subsubsection{Embedding of the SM gauge group}

Let us be more concrete regarding the embedding of the SM gauge group into the global symmetry group. A simple choice for the number of flavors and colors of the electric theory corresponds to~\cite{Craig:2009hf}
$$
N_f= 12\;,\;\t N_c =1 \;\Rightarrow\; N_c=23\,.
$$
Recalling that the $SU(N_f)$ global symmetry is broken to $SU(N_f-\t N_c)$, the embedding of $G_{SM}$ is given by
\begin{equation}\label{eq:embed}
Q \, \sim \, ({\bf 5} + {\bf \bar{5}} + {\bf 1}) + {\bf 1}\;,
\t Q \, \sim \, ({\bf \bar{5}} + {\bf 5} + {\bf 1}) + {\bf 1}\,.
\end{equation}
Actually, a realistic theory of flavor with inverted hierarchies requires larger $N_f$ and $N_c$; we refer the reader to~\cite{Craig:2009hf} for details.

Decomposing the mesons as in Eq.~(\ref{eq:paramPhi}),
\begin{equation}\label{decompPhi1Phi2}
\Phi_i= \left(\begin{matrix} Y_{i,\,1 \times 1} & Z^T_{i,\,1 \times 11} \\ \tilde Z_{i,\,11 \times 1} &X_{i,\,11 \times 11}\end{matrix} \right)\;,\;i=1,2\,,
\end{equation}
the SM quantum numbers of the $X$ fields are
\begin{equation}\label{eq:reps2}
X_i \sim ({\bf 10}+ \bar{\bf 5}) +  \bar{\bf 5}+\left[2 \times {\bf 24} + {\bf 15} + {\bf \overline{15}} + {\bf \overline{10}} + 2 \times {\bf 5} + 3\times {\bf 1} 
\right]\,.
\end{equation}
Each of the mesons $X_1$ and $X_2$ yields one composite SM generation ${\bf 10}+ \bar{\bf 5}$, while $H_d$ comes from the extra $\bar{\bf 5}$ in $X_1$.

Except for the singlets, the representations ${\bf R}$ inside the square brackets give rise to extra matter that has to be removed from the low energy spectrum. This is done by introducing spectator fields $S_{\bf \bar{R}}$ with couplings to the unwanted matter
\begin{equation}\label{eq:additionalMatter}
W_{el} \supset  \lambda_{1{\bf R}} \sum_{\bf{R}} S_{1 {\bf{\bar{R}}}} (Q \t Q)_{\bf R} +
\lambda_{2{\bf R}} \f{1}{\Lambda_0} \sum_{\bf{R}} S_{2 {\bf{\bar{R}}}} (Q U \t Q)_{\bf R} \,.
\end{equation}
In the IR this gives masses of order $\Lambda$ and $\Lambda^2/\Lambda_0$ to the unwanted matter.
The same procedure is used to lift the linear combination of $(Z_i, \t Z_i)$ that does not couple to the supersymmetry breaking field (\ref{eq:Phidef}).

Importantly, after turning on the chiral deformation (\ref{eq:additionalMatter}), the composite SM fermions only acquire masses from the Yukawa couplings. The $\muphi$ perturbations produce a negligible admixture with spectators.

\subsubsection{Yukawa couplings}\label{subsubsec:Yukawas}

We assume that at some high scale $M_{\rm flavor}> \Lambda$ there is new physics responsible for generating couplings between the Higgs and the gauge invariant mesons~\cite{Franco:2009wf, Craig:2009hf},
\begin{eqnarray}\label{eq:Yuk}
W_{y_u} & \sim & \frac{1}{M_{\rm flavor}^4} (QU\t Q) H_u (QU\t Q) + \frac{1}{M_{\rm flavor}^3} (Q\t Q) H_u (QU \t Q) + \f{1}{M_{\rm flavor}^2} (Q\t Q) H_u (Q\t Q)+\nonumber \\
& &  + \f{1}{M_{\rm flavor}} (Q\t Q) H_u \Psi_3 + \f{1}{M_{\rm flavor}^2} (Q\t U Q) H_u \Psi_3 +\Psi_3 H_u \Psi_3\,.
\end{eqnarray}
Notice that only couplings to the SM fields are generated, because the extra matter in the mesons is lifted using (\ref{eq:additionalMatter}).

After the theory confines, these irrelevant interactions become Yukawa couplings in terms of the elementary mesons,
\begin{eqnarray}
W_{y_u} &\sim&  \frac{\Lambda^4}{M_{\rm flavor}^4} \Phi_2 H_u \Phi_2 + \frac{\Lambda^3}{M_{\rm flavor}^3} \Phi_1 H_u \Phi_2 +
 \f{\Lambda^2}{M_{\rm flavor}^2} \Phi_1 H_u \Phi_1 +\nonumber\\
&+& \f{\Lambda}{M_{\rm flavor}} \Phi_1 H_u \Psi_3 + \f{\Lambda^2}{M_{\rm flavor}^2} \Phi_2 H_u \Psi_3 +\Psi_3 H_u \Psi_3. 
\end{eqnarray}
In terms of
\begin{equation}
\epsilon \equiv \frac{ \Lambda}{M_{\rm flavor}}
\end{equation}
the up-type fermion Yukawa matrix becomes
\begin{equation}\label{eq:Yukawatext}
y_U \sim \left( \begin{array}{ccc}
\epsilon^4 & \epsilon^3 & \epsilon^2 \\
\epsilon^3 &\epsilon^2 & \epsilon      \\
\epsilon^2 & \epsilon    & 1 
\end{array}
\right)\,
\end{equation}
(up to order one numbers). The correct flavor texture is generated for $\epsilon \sim 10^{-1} - 10^{-2}$, so that the ``flavor'' scale is approximately one decade above the compositeness scale $\Lambda$.

On the other hand, the down- and lepton- type Yukawa couplings are generated from couplings to the composite Higgs $H_d \subset (Q \t Q) / \Lambda$. For simplicity, it is assumed that such couplings are also generated at the scale $M_{\rm flavor}$ : \footnote{For instance, in models that unify one has $\Lambda \sim M_{GUT}$~\cite{Craig:2009hf}, and then it's natural to assume that all the irrelevant interactions are produced near the Planck scale.}
\begin{eqnarray}\label{eq:Yuk2}
W_{y_d, y_l} & \sim & \frac{1}{M_{\rm flavor}^5} (QU\t Q) (Q \t Q) (QU\t Q) + \f{1}{M_{\rm flavor}^3} (Q\t Q)^3+ \f{1}{M_{\rm flavor}}\Psi_3 (Q \t Q) \Psi_3 + \ldots \nonumber \\
& \to & \epsilon^5  \Phi_2 H_d \Phi_2 + \epsilon^3 \Phi_1 H_d \Phi_1 + \epsilon \Psi_3 H_d \Psi_3 + \ldots
\end{eqnarray}
This gives
\begin{equation}\label{eq:ydl}
y_d \sim y_l \sim \epsilon \,y_u \,.
\end{equation}

In our proposal, this result explains dynamically the smallness of the bottom and tau mass compared to the top mass. However, the hierarchy of second generation masses is smaller, and in the first generation $m_d$ is actually a bit heavier than $m_u$. Eq.~(\ref{eq:ydl}) at large $\tan \,\beta$ would produce too light fermion masses in the first two generations, unless additional structure is present in the Yukawa matrices. The realistic models we discuss below have $\tan\, \beta \sim 6 - 10$ and then this is not a serious issue because factors of 5 or so in the Yukawa matrices are enough to produce realistic fermion masses. On the other hand, for parametrically large $\tan \,\beta$, another mechanism for generating the down type masses of the first two generations would be needed. See discussion below (\ref{eq:Ad}).

\subsection{New vacua and lifetime}\label{subsec:newvacua}

The example (\ref{eq:reps2}) considered here requires spectators in vector-like representations, with superpotential interaction (\ref{eq:additionalMatter}). This leads to new metastable vacua near the origin, where the spectators and $\rho$ fields acquire nonzero expectation values.\footnote{These vacua were first found by D. Green, A. Katz and Z. Komargodski.} The vacuum energy of the metastable configurations depends on the eigenvalues of $\hat \mu^2$. 

For a spectator in a vector-like representation coupling to a diagonal block of the mesons of dimension $n$, the new metastable configuration has~\cite{Craig:2010ip}
\begin{eqnarray}\label{eq:newVEV}
\langle \chi^T \chi \rangle &=& \mu_1^2\,\mathbf 1_{\t N_c-1}\;,\;\langle \rho^T \rho \rangle = \mu_2^2\,{\rm diag} (0,\ldots,0,n) \nonumber\\
\langle S \rangle &=& \frac{\mu_2^2}{\lambda \Lambda}\,{\rm diag}(1,\ldots,1,-(n-1)) \,,
\end{eqnarray}
after an appropriate flavor rotation. The vacuum energy reads
\begin{equation}\label{eq:V0new}
V_0 = h^2 \mu_1^4+2(N_f-\t N_c-n)\, h^2 \mu_2^4\,.
\end{equation}
The energy of the new vacua is higher than the ISS one for
\begin{equation}
\mu_2^2 \le \frac{1}{n^{1/2}} \mu_1^2\,.
\end{equation}
In this case our metastable solution would be stable against decay towards (\ref{eq:newVEV}).

However, for the matter content (\ref{eq:reps2}) this would require $\mu_i$'s differing by an order of magnitude. Such a choice would in turn lead to a strong suppression in the fermion masses (\ref{eq:mgaugino}), (\ref{eq:mu}) and a split spectrum. In order to avoid this, in what follows we restrict to $\mu_1^2 = \mu_2^2 \equiv \hat \mu^2$. Then we have to make sure that the decay rate to the new vacua (\ref{eq:newVEV}) is small enough. 

We have checked this by performing an explicit numerical evaluation of the action of the multi-field semiclassical tunneling configuration, or ``bounce'' \cite{Coleman:1977py}, associated with  the decay to a vacuum involving nonzero VEVs for a spectator $S_{\bf 25}$, corresponding to $n=5$ in eq.~\eqref{eq:newVEV}. To compute the bounce, we have assumed that it only involves nonzero profiles for the real part of the fields $\rho,\t\rho,\chi,\t\chi,S$, as well as (taking for simplicity $\t N_c=1$, though the results apply equally to $\t N_c>1$), $\rho=\tilde\rho\equiv(0,0,0,0,r),\,\chi=\t\chi\equiv x,\,S\equiv{\rm diag}(s,s,s,s,-4s)\,$. In the $\mu_\phi=0$ case, this can be justified from the equations for the bounce and the boundary conditions; when considering $\mu_\phi\neq0$, one should also consider tunneling along the $X$, $Y$ field directions, which should yield smaller decay rates. The bounce configuration that we considered involves thus three fields, $x,r,s$; in order to obtain it numerically we have used the technique of ref.~\cite{Konstandin:2006nd}, taking as potential the tree-level one plus one-loop contributions from the gauge fields, which we modelled as
\begin{align}
 V_g\sim\frac{g^2}{16\pi^2}|\rho+\t\rho^*|^2|\chi+\tilde\chi^*|^2.
\end{align}
The above choice is motivated as follows: around each vacua with either $\langle\chi\rangle$ or $\langle\rho\rangle$ nonzero, it generates mass-terms %
for either the $\rho$ or $\chi$ fields -which in particular stabilize the $\rho+\rho^*$ pseudo-Goldstone direction in the ISS vacuum, ensuring that there actually is a potential barrier- but the energies of the vacua are not affected, as required by the fact that on each of the minima the SUSY vector multiplets decouple from the SUSY breaking, at least to lowest order.

\begin{figure}[h]
\begin{center}
\epsfig{file=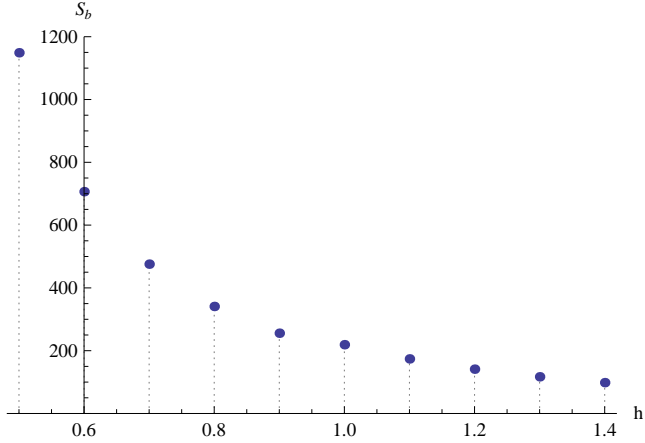}
\caption{Bounce action as a function of $h$, for  $\kappa_\Lambda=385$.} 
\label{fig:bounceaction}
\end{center}
\end{figure}

Ignoring $\mu_\phi$, it can be easily seen that the action of the bounce only depends on the dimensionless parameters $h$ and $\kappa_\Lambda\equiv\Lambda/\hat\mu$. The dependence on $\kappa_\Lambda$ turns out to be very weak, which is a consequence of the small VEV of the field $s$ in the new vacuum, which is suppressed by $\Lambda/\hat\mu$, see eq.~\eqref{eq:newVEV}. Hence the lifetime of the ISS vacuum depends mainly on $h$. We have seen that the bounce action $S_b$ goes like inverse power laws of $h$, so that the lifetime $\Gamma\sim\exp{S_b}$ increases very rapidly for decreasing $h$. This can be understood as a consequence of the fact that $h$ controls the energy difference between the two vacua. The explicit computations show that a bounce action $S_b\gtrsim400$, which is enough to guarantee a lifetime greater than the age of the Universe, is attainable for $h\lesssim0.75$ (see Fig.~\ref{fig:bounceaction}). For more details about the computation, we refer the reader to a future note \cite{inprep}.

\subsection{Summary of the model}\label{subsec:summary}

We end this section by summarizing the model and its interactions, together with the effects of integrating out the composite messengers. 

First, the scales of the model are the following. The dynamical scale is $\Lambda$, below which we switch to the weakly coupled magnetic description; in examples with unification this is $\Lambda \sim M_{GUT}$ but otherwise it can be smaller. There are two scales above $\Lambda$; first, $M_{\rm flavor}$ is the scale at which the interactions between the Higgs and the other SQCD mesons are generated. And $\Lambda_0$, introduced in (\ref{eq:DelWU}), controls the irrelevant polynomial deformations in the mesons --these are responsible, in particular, for breaking the R-symmetry. These two scales are taken to be roughly of the same order of magnitude, and are one or two orders of magnitude above $\Lambda$. The other dimensionful parameter of the model is the electric quark mass $m$; it sets the scale of supersymmetry breaking $\hat \mu^2 \sim m \Lambda$, and $m/\Lambda \ll 1$ is required for metastability. Actually, this mass parameter is not strictly required because supersymmetry breaking can also be obtained from the marginal deformation $\Delta W_{el} \sim Q U \t Q$ in (\ref{eq:DelWU}).

The IR interactions are given by
\begin{equation}\label{eq:Wfullmag}
W_{mag}= W_{O'R}+ \lambda_u \tr(Z H_u \t Z) + \tr\,\rho\left( \lambda_d  H_d +h_1 \Phi_1^{SM}+ h_2 \Phi_2^{SM}\right)\t \rho  + W_{Yuk}\,.
\end{equation}
Let's describe the different contributions to $W_{mag}$. The O'Raifeartaigh-type terms are, from Eq.~(\ref{eq:WmagU2}),
\begin{equation}
W_{O'R} = - h\hat \mu^2 \, \tr\, X + h \tr(\rho X \t \rho)+ \frac{1}{2}h^2 \muphi\,\tr X^2+ h \hat \mu \tr (Z \t \rho+ \t Z \rho)
\end{equation}
where $X$ arises from the $(N_f - \t N_c) \times (N_f - \t N_c)$ sub-block in the linear combination (\ref{eq:Phidef}); $(Z, \t Z)$ also originate from this combination, albeit from the $1 \times (N_f - \t N_c)$ block. The fields $(Y, \chi, \t \chi)$ have not been included; they have a supersymmetric spectrum and don't contribute at one loop. Also, recall that we have restricted to coincident linear terms.

Next we have the interaction between $H_u$ and the $Z$ messengers; it comes from a dimension 5 operator in the UV so that $\lambda_u \ll 1$. The couplings between the $\rho$ messengers and the composite MSSM fields originate from (\ref{eq:WmagU}). $H_d$ and the first generation arise from the dimension two meson, so they have the same coupling $h_1=\lambda_d$. The second generation $\Phi_2^{SM}$ has a different coupling, denoted by $h_2$. Both $\lambda_d$ and $h_2$ are of the order of $h$, while $\lambda_u$ is parametrically smaller; this is due to the fact that $H_d$ and $\Phi_i^{SM}$ are composites, while $H_u$ is elementary.

The effective theory at the supersymmetry breaking scale $h \hat \mu$ is obtained by integrating out $(\rho, Z)$. The dominant effects arise at one loop (CW potential) and two loops (gauge mediated contributions),
\begin{equation}\label{eq:Veff}
V_{eff}= V_{tree}+\frac{1}{64 \pi^2}\,{\rm Str} \;\mc M^4\,\left(\log \frac{\mc M^2}{(h \hat \mu)^2}-\frac{3}{2}\right)+ V_{GM}^{2-loop}\,.
\end{equation}
While we discuss these calculations in detail in Appendix \ref{app:one-loop}, let us analyze qualitatively the various soft terms that are produced. There are three types of effects:
\begin{enumerate}
\item [a)] CW contributions: they affect scalars that have tree level couplings to the messengers,
\begin{equation}\label{eq:mcomposite}
m_{CW}^2 \sim \frac{h^2\t N_c}{16 \pi^2} (h \hat \mu)^2 \left(1 + \mathcal O \left(\frac{\langle X \rangle}{\hat \mu} \right)\right)\,.
\end{equation}
The corrections proportional to $\langle X \rangle / \hat \mu$, coming from the breaking of R-symmetry, are important for some of the soft parameters, and are taken into account in the explicit analysis of \S \ref{sec:pheno}.
\item [b)] Elementary sfermion masses come predominantly from  two loop gauge mediated effects,\footnote{These two loop computations for the model of~\cite{Essig:2007xk}, which our work uses, have been analyzed by R.~Essig and J.~F.~Fortin~\cite{private}. We thank them for their help.}
\begin{equation}\label{eq:melementary}
m_{GM}^2 \sim \t N_c\left(\frac{g^2}{16 \pi^2}\right)^2 (h \hat \mu )^2\,.
\end{equation}
\item [c)] Gaugino and higgsino masses appear at one loop and are proportional to the R-symmetry breaking parameter (for small $X/\hat \mu$)
\begin{equation}\label{eq:mfermion}
m_{\psi} \sim \frac{\t N_c}{16 \pi^2} \langle h X \rangle  \sim \muphi\,.
\end{equation}
\end{enumerate}
The soft parameters for the Higgs sector produced by the CW potential were given in \S \ref{subsec:EW}. Gauge-mediated contributions to the Higgs potential (and the nonzero VEV $\langle h X \rangle$) have to be taken into account as well.

We see that integrating out the heavy messengers produces masses squared (\ref{eq:mcomposite}) for the composite MSSM fields that are one loop below the supersymmetry breaking scale. On the other hand, elementary fields get their masses predominantly from gauge mediated effects (\ref{eq:melementary}), so they are two loops below the supersymmetry breaking scale. Finally, the gauginos and higgsinos end up having masses proportional to $\muphi$ (recall that $\langle h X \rangle \sim 8 \pi^2 \muphi$ so that the loop factors cancel in the fermion masses). In practice, the higgsinos tend to be quite light because their mass receives an extra suppression proportional to $\lambda_u \lambda_d$. This spectrum is of the type considered for instance in~\cite{Cohen:1996vb}, where the first two generation sfermions (plus in our case $H_d$) are decoupled to the multi-TeV range, while around the TeV scale one only has third generation matter, gauginos and a light Higgs. 

Requiring the masses of gauginos and third generation sfermions to be at around $1\;{\rm TeV}$ sets, for $\t N_c={\cal O}(1)$,
\begin{equation}
h \hat \mu \sim 200\;{\rm TeV}\;,\;\muphi \sim 1\;{\rm TeV}\;\Rightarrow\;x \lesssim 1\,,
\end{equation}
where the relation between $\muphi$ and $X/\hat \mu$ is given, to lowest order, in Eq.~(\ref{eq:Xvev}). In this case, the composite masses are of the order
\begin{equation}
m_{H_d} \sim m_{\t Q_i} \sim 10 - 20\;{\rm TeV}\,.
\end{equation}
In order to get higgsinos around $100 - 200\, {\rm GeV}$, we take
\begin{equation}
\lambda_u \sim 0.1\;,\;\lambda_d \sim h \sim h_i \sim 1\,.
\end{equation}
This is the parameter range of interest in what follows.\footnote{It is necessary to point out that the simple mechanism for EWSB discussed in \S \ref{subsec:EW}, where the messengers generate a tachyonic mass for $H_u$, receives now important modifications. The reason is that having large enough gaugino masses and $\mu$ term requires $x \sim 1$ and around these values, the one loop contribution to $m_{H_u}^2$ (first term in Eq.~(\ref{eq:mHu-param})) transitions from negative to positive. This has to be added to a positive two loop mass, so that the total mass for $H_u$ is positive in the regime of interest. As explained below, the breaking of $SU(2) \times U(1)$ will be produced by a combination of dynamical and radiative effects.} Parameter ranges, spectra and other phenomenological properties are discussed in \S \ref{sec:eft} and \S \ref{sec:pheno}.

Mixing between the Higgs fields and the supersymmetry breaking sector produces one loop A-terms for the composite generations,
\begin{equation}
L_{A-terms}=  A_d\, \t Q H_d\; {\t {\bar d}}  +A_l\, \t L H_d {\t {\bar e}}  +A'_d\, \t Q H_u^\dag {\t {\bar d}} + A_l'\, \t LH_u^\dag {\t {\bar e}}\,.
\end{equation}
Nonzero $A_u$ terms would require messengers transforming in a ${\bf 10}+\overline{\bf 10}$ representation.
To lowest order in the R-symmetry breaking parameter, the A-terms are found to be
\begin{eqnarray}\label{eq:Ad}
A_d&=& \frac{\lambda_d \t N_c h_i^2}{8 \pi^2}\,\left(\frac{41}{15}-4 \log 2\right) \frac{\langle  X \rangle^3}{\hat \mu^3}\,h \hat \mu+ \ldots \nonumber\\
A_d'&=& \frac{\lambda_u \t N_c h_i^2}{8 \pi^2}\, \left(\frac{129}{20}-9 \log 2\right) \frac{\langle  X \rangle^3}{\hat \mu^3}\,h \hat \mu + \ldots \,,
\end{eqnarray}
The contribution to the third generation is negligible, as in usual gauge mediation. It is possible to choose the SM embedding so that the A-terms are diagonal in generation space; this again requires $N_f$ and $N_c$ larger than the minimal (\ref{eq:embed}).

Due to the cubic power of $X/\hat \mu$ (and the $\lambda_u$ factor in the case of $A_d'$), these A-terms are parametrically smaller than the other CW soft terms. However, they may still give interesting low energy effects, particularly the $A'$ term, which is rarely included. In particular, a one loop diagram involving a bino and an $A_d'$ insertion generates a fermion mass operator
$$
L_f = - y_d'\, Q H_u^* \bar d\,.
$$
This mechanism for generating quark masses radiatively could become useful in regions of very large $\tan \beta$. For a recent work on other loop induced fermion masses and references see~\cite{Dobrescu:2010mk}.

Let us end our analysis of the single sector model with a brief discussion of flavor changing neutral currents (FCNCs). These are produced because the fermion mass matrix is not diagonal in the same basis as the sfermion mass matrix and A-terms. After changing to the fermion mass eigenbasis, the sfermion mass matrix acquires off-diagonal components that, through loop effects, can induce for instance $K^0 - \bar K^0$ mixing or  rare decyas like $\mu \to e \gamma$. In our model the soft masses in the interaction basis are diagonal in generation space, so the off-diagonal components induced by the above rotation are much smaller than the diagonal elements. In this case, constraints from FCNCs place upper bounds on ratios of the off-diagonal mass components divided by the average sfermion mass~\cite{Gabbiani:1996hi}.

Soft masses (\ref{eq:mcomposite}) give contributions $\sim m_{\t Q}^2 (\t Q^*_L \t Q_L + \t Q^*_R \t Q_R)$ which do not change chirality. For these elements, the strongest constraint comes from $K-K$ mixing. These effects were studied in~\cite{Craig:2009hf}, where it was shown that the bounds are satisfied for composite masses around $10-20\;{\rm TeV}$. The analysis is similar in our case.

On the other hand, after setting the Higgs to its VEV, A-terms give mass-contributions $A_d' v_u (\t Q^*_L \t Q_R + \t Q^*_R \t Q_L)$ that change chirality. These lead to potentially new sources of flavor violation. We find that the strongest constraint comes from the lepton flavor violating decay $\mu \to e \gamma$. This bound is satisfied in our model with masses at $10-20\;{\rm TeV}$, if we require some degeneracy (at the $10-20\,\%$ level) between the soft masses. This can be obtained by a mild tuning of the electric theory parameters, and the tuning decreases with increasing masses. In summary, the model leads to flavor changing effects satisfying the experimental bounds, by a combination of decoupling, diagonal soft terms in the interaction basis, and a mild degeneracy between CW masses.


\section{Effective field theory analysis}\label{sec:eft}

In \S \ref{sec:compositeH}--\S \ref{sec:ssector} we have described the RG evolution of the system, starting from the microscopic SQCD theory, the generation of flavor textures at $M_{{\rm flavor}}$ by dimensional hierarchy, and then the appearance of light composites giving rise to MSSM matter below the compositeness scale $\Lambda$. At the scale $h \hat \mu$ the strongly coupled sector admits an effective description in terms of Eq.~(\ref{eq:Wfullmag}). Supersymmetry is dynamically broken and integrating out the heavy messengers via the Coleman-Weinberg potential (plus gauge mediated effects) generates finite soft terms that were summarized in \S \ref{subsec:summary}.

The aim of the next two sections is to link the physics taking place at the messenger scale with that at $M_Z$, putting special emphasis on the scalar potential for the Higgs fields and the breaking of the electroweak symmetry.  A careful analysis is motivated by the fact that models with inverted hierarchies and a hybrid Higgs sector exhibit quite different properties from the usual MSSM spectrum. Since this discussion will be somewhat technical, the reader interested mainly in the phenomenology at the TeV scale can move directly to \S \ref{sec:pheno}.

Since so far we described the dynamics of the strongly coupled sector using Seiberg duality (which follows from a Wilsonian analysis of the interactions) it is natural to study the flow towards the EW scale in a physical approach that, as the Wilsonian one, takes into account the decoupling of energy scales. We proceed by constructing successive effective theories, integrating out particles at their thresholds; this should be done using a mass-dependent renormalization scheme, which we approximate by defining the $\beta$-functions in a piecewise manner at each energy interval. Besides quantum corrections coming from the RG flow, we will take into account finite CW effects and two-loop loop gauge mediated masses.

As explained in \S \ref{subsec:summary}, below the messenger scale there are three mass hierarchies, corresponding to the composite fields (see (\ref{eq:mcomposite})), third generation sfermions (given in~\eqref{eq:melementary}), and gauginos (Eq.~\eqref{eq:mfermion}). The discussion is simplified by integrating out simultaneously particles with similar masses, resulting in only three thresholds below the messenger scale. We do not decouple the third generation sleptons since their effect in the RG running of the other parameters is negligible, and in some ranges they tend to be quite light.

In the following we will focus on the RG evolution produced by these mass hierarchies, without giving specific numerical details. We will identify the relevant degrees of freedom in the different energy regimes and determine the dependence of the soft masses and Higgs VEV at the $M_Z$ scale on the messenger scale parameters $(h \hat \mu, \muphi)$ and couplings $(h_i,\lambda_d, \lambda_u)$. More detailed numerical results, the resulting spectra and low energy phenomenology are presented in \S \ref{sec:pheno}.

\subsection{Parametrization of soft terms}\label{subsec:param}

It is useful to first recast the soft parameters at the messenger scale in a way that makes manifest their dependence on $(h \hat \mu, \muphi, h_i,\lambda_i)$. Throughout, we make use of the running scale 
$$
t=\log \,\frac{Q}{h \hat \mu}\,.
$$
It was argued in \S \ref{subsec:mu} that the spontaneous R-symmetry breaking $\langle h X \rangle$ dominates over the explicit breaking $\muphi$, so we find it convenient to trade $\muphi$ for the dimensionless combination
\begin{equation}
x \equiv \frac{\langle h X \rangle}{h \hat \mu}\,.
\end{equation}

As explained in Appendix \ref{appsubsec:soft}, the dependence of the  masses of composite particles on the microscopic parameters is given by
\begin{eqnarray}\label{eq:mcomp-param}
m^2_{H_d}(0) &=& \left[\frac{\lambda_d^2}{8 \pi^2} \hat V^{cw}_{H_d H^*_d}(h,x)+ \sum_a C_a^{H_d} \left(\frac{g_a^2(0)}{16 \pi^2}\right)^2\,f_{gm}(h,x)\right](h \hat \mu)^2\nonumber\\
m^2_{\t Q_i}(0)&=&\left[\frac{h_i^2}{8 \pi^2} \hat V^{cw}_{\tilde Q_i \tilde Q^*_i}(h,x)+ \sum_a C_a^{\t Q} \left(\frac{g_a^2(0)}{16 \pi^2}\right)^2\,f_{gm}(h,x)\right](h \hat \mu)^2\;,\;i=1,\,2\,.
\end{eqnarray}
Here $\hat V^{cw}_{H_d H^*_d}= \hat V^{cw}_{\tilde Q_i \tilde Q^*_i}$ are second derivatives of a dimensionless potential $\hat V^{cw}$ defined in terms of $V_{CW}$ in Eq.~(\ref{eq:defVhat}). This is useful in order to show explicitly the loop factors and dependence on $(h\hat \mu)$. Similarly, $f_{gm}$ comes from the gauge-mediated two loop potential (third term in Eq.~(\ref{eq:Veff})).

In the limit of small $x$, $\hat V^{cw}_{H_d H^*_d}\sim\t N_c(\log 4 -1$) as given in (\ref{eq:softmasses}); the full $x$ dependence is taken into account in the numerical results of \S \ref{sec:pheno}. For these masses, one loop contributions dominate over gauge-mediated two loop effects, which will be neglected in the analytical formulae of this section.

Third generation sfermion masses arise at two loops,
\begin{equation}\label{eq:melem-param}
m^2_{\t Q_3}(0) = \sum_a C_a^{\t Q} \left(\frac{g_a^2(0)}{16 \pi^2}\right)^2\,f_{gm}(h,x)\,(h \hat \mu)^2\,.
\end{equation}
On the other hand, $m_{H_u}$ receives both one loop (from the trilinear coupling $W \supset \lambda_u Z H_u \t Z$) and two loop contributions, 
\begin{equation}\label{eq:mHu-param}
m^2_{H_u}(0) = \left[\frac{\lambda_u^2}{8 \pi^2} \hat V^{cw}_{H_u H^*_u}(h,x)+ \sum_a C_a^{H_u} \left(\frac{g_a^2(0)}{16 \pi^2}\right)^2\,f_{gm}(h,x)\right](h \hat \mu)^2\,.
\end{equation}
Even though they appear at different loop orders, both contributions are comparable because of the smallness of $\lambda_u$. The $B_\mu$ term comes from mixed derivatives of the CW potential,
\begin{equation}
B_\mu(0)= -\frac{\lambda_u \lambda_d}{8 \pi^2}\, \hat V^{cw}_{H_u H_d}(h,x)\,(h \hat \mu)^2\,.
\end{equation}
From Eq.~(\ref{eq:softmasses}), the small $x$ behavior is $ \hat V^{cw}_{H_u H_d}\sim -\t N_c(1-\log 2)$.

Finally, gaugino and higgsino masses are parametrized as
\begin{equation}\label{eq:mferm-param}
m_{\lambda_a}(0)= \frac{g_a^2(0)}{16 \pi^2} \left(x f_\lambda(h,x) \right)\,h\hat \mu\;\,,\,\;\mu(0) = \frac{\lambda_u \lambda_d}{16 \pi^2} \left(x f_\mu(h,x) \right)\,h\hat \mu\,.
\end{equation}
As discussed in \S\ref{subsec:mu}, these masses vanish linearly with the R-symmetry breaking parameter $x$, and
$f_\lambda$ and $f_\mu$ have order one values at $x=0$. Choosing the microscopic parameters to be real, there are no phases in the soft masses.

\subsection{Flowing down to the top scale: effective potential method}\label{subsec:method}

The flow from the messenger scale down to the top scale is done as follows. Below the messenger scale the effective theory is the MSSM, with boundary values for the soft parameters obtained in the previous subsection. Between $h \hat \mu \sim 200\;{\rm TeV}$ and $m_{CW} \sim 10-20\;{\rm TeV}$ the evolution is dictated by the one loop MSSM RG equations.

At energies $Q \sim m_{CW}$, $H_d$ and the heavy generations are integrated out by absorbing into tree-level parameters their contributions to the one loop potential
\begin{equation}\label{eq:VCWgen}
V_{CW}=\frac{1}{64 \pi^2}\,{\rm Str} \;\mc M^4\,\left(\log \frac{\mc M^2}{Q^2}-\frac{3}{2}\right)\,,
\end{equation}
where now the role of the cutoff of the effective theory that results from integrating out the heavy particles is played by the threshold scale $Q \sim m_{CW}$, and $\mc M$ is the $H_u$-dependent mass-matrix for the heavy MSSM particles, evaluated at $Q$. These contributions are absorbed into the low energy masses, producing finite corrections of order
$$
\sim\,\frac{m_{CW}^2}{16 \pi^2}\,.
$$
Such contributions are similar to the effects from integrating out the messengers, with the replacement $h \hat \mu \to m_{CW}$. Indeed, in composite models with inverted hierarchies the heavy generations behave much like messengers (with the obvious difference that the fermions are kept in the low energy theory). 

The outcome is a new effective theory with cutoff $m_{CW}$, containing only the third generation sfermions, gauginos and $H_u$, plus the SM matter. In order to implement correctly the decoupling of the heavy particles, the $\beta$-functions have to be computed in a mass-dependent scheme~\cite{Appelquist:1974tg}. As is usually done, we will approximate the mass-dependent $\beta$-functions by defining them in a stepwise fashion, starting from their MSSM values in a mass independent scheme and implementing the decoupling of massive particles by simply removing their contributions to the RG equations below each threshold. 
 This procedure is then repeated for the stop/sbottom and heavy gauginos. 
 
 In the last stage, EWSB is computed in a theory at the top scale, where the dominant effects come from $H_u$ and the top quark.\footnote{While these are one loop effects in the low energy theory, they arise at two loops in the theory (\ref{eq:Wfullmag}) containing all the fields of the magnetic theory. For $h \lesssim 1$ and $\lambda_u \ll 1$, such effects dominate over the two loop contributions involving the messengers only, which we have hence neglected. Two loop effects in the ISS context have been discussed in~\cite{Giveon:2008wp}.} For the purpose of obtaining approximate analytical formulae, the gauge and Yukawa couplings are considered as inputs at the high scale.\footnote{The parameters that are fixed by low energy experiments (such as gauge couplings) have to be evolved towards the messenger scale accross different thresholds, whose scales depend in turn on the high energy values of these couplings. This is solved iteratively in the results presented in \S \ref{sec:pheno}.} The different energy ranges are analyzed, in turn, in \S \ref{subsec:comp-scale} and \S \ref{subsec:top-scale}. 

To make the discussion more concrete, consider a complex scalar $\phi$ with soft  mass $m_{\phi}^2$, and a marginal coupling to the light Higgs,
\begin{equation}
V =  \left(m_{\phi}^2+ y^2 |H|^2 \right) |\phi|^2+ \ldots
\end{equation}
For energy scales above  $m_{\phi}^2$, the running Higgs mass $m_{H_u}^2(t)$ obeys, in a mass-independent scheme,
\begin{equation}
\frac{d m_{H_u}^2}{dt}= \frac{1}{8 \pi^2}\,m_{\phi}^2 + \ldots
\end{equation}
Below the scale $Q=m_{\phi}$, in order to approximate the change of the $\beta$-function in a mass-dependent scheme, this contribution is set to zero, and the particle is integrated out using (\ref{eq:VCWgen}), which now reads
\begin{equation}\label{eq:toyCW}
V_{CW}(Q^2=m_\phi^2) = -\frac{m_{\phi}^2}{16 \pi^2}\,y^2 |H|^2+ \mc O(|H|^6/m_\phi^2)\,.
\end{equation}
The quartic term vanishes at the threshold. Nonrenormalizable terms (kept frozen at and below the thresholds) give negligible contributions, being suppressed by inverse powers of $m_\phi$. They are ignored in the analytic results below.

Using first the RGEs valid at
$$
t > t_\phi = \log \frac{m_\phi}{h \hat \mu}
$$
to compute $m_{H_u}^2(t_\phi)$ --with the boundary condition~(\ref{eq:mHu-param})-- and then combining this with Eq.~(\ref{eq:toyCW}), gives a Higgs soft mass $\overline m_{H_u}$ in the effective theory below the threshold  at $t_\phi$ equal to
\begin{equation}\label{eq:toy1}
\overline m^2_{H_u}(t_\phi)=m_{H_u}^2(t_\phi) -\frac{m_{\phi}^2}{16 \pi^2}\,y^2\,.
\end{equation}
For $t < t_\phi$, the contributions of $m_\phi^2$ to the $\beta$-functions are set to zero and, as shown in~\cite{Carena:1995wu}, the $\beta$-function for the Higgs quartic coupling $L \supset - \frac{\lambda}{2} |H_u|^4$ receives a correction
\begin{equation}\label{eq:beta-lambda}
\frac{d \lambda}{dt}=-\frac{y^4}{16 \pi^2} + \ldots
\end{equation}
This plays an important role in increasing the mass of the physical light neutral Higgs.

As a technical aside, it should be noted that this procedure of decoupling of particles can be justified from a diagrammatic interpretation of the Coleman-Weinberg potential, making use of the decoupling theorem of~\cite{Appelquist:1974tg} to argue that the particle in question can be eliminated from the effective theory (and $\beta$-functions) below its threshold. A transition to a mass-dependent scheme has to be made so that the theorem applies. This is implemented, in the procedure explained above, by absorbing at each threshold the quantum contributions due to the corresponding particle into the tree-level parameters of the effective potential. As a consequence, such parameters suffer threshold corrections and have a non-continuous dependence on the renormalization scale $Q$. There are, however, no discontinuities in the scale dependence of the full effective potential.

The effective potential method for EWSB computations has been widely used (see e.g.~\cite{Carena:1995wu}). Combining the RG running of parameters with the CW potential  minimizes the dependence of the effective action on the renormalization scale~\cite{Gamberini:1989jw}. Also, the implementation of the decoupling of particles avoids the breakdown of perturbation theory due to the appearance of large logarithms in a mass-independent renormalization scheme, and the integration of fields at each threshold implements some higher loop corrections. Typically, this approach is applied to the top/stop sector, because the other generations give negligible contributions. Our main point here is that, in models with inverted hierarchies, effects from the heavy composites are also important in the calculation of the EWSB vacuum, and have to be consistently taken into account.

\subsection{EFT at the scale of the composite MSSM fields}\label{subsec:comp-scale}

Let us describe the RG evolution from $h \hat \mu$ to $m_{CW}$, the scale of the MSSM composites. The soft parameters at $t=0$ were defined before and, as usual, the effective potential in the Higgs sector is
\begin{eqnarray}\label{eq:fullVH}
V_{eff}&=&m_{H_u}^2 |H_u|^2 + m_{H_d}^2 |H_d|^2-B_\mu (H_u H_d + c.c.)+\frac{g_2^2}{2}|H_u^+ H_d^{0*}+H_u^0 H_d^{-*}|^2+\nonumber \\
&+&\frac{1}{8}(g_2^2+ \frac{3}{5}g_1^2)\left(|H_u^0|^2-|H_d^0|^2+|H_u^+|^2-|H_d^-|^2 \right)^2+V^{MSSM}_{CW}+\ldots
\end{eqnarray}
Here `$\ldots$' includes irrelevant operators suppressed by inverse powers of $h \hat \mu$ which can be safely ignored, and corrections from the microscopic theory --they are negligible by the arguments given in~\cite{Intriligator:2006dd}. It is convenient to introduce the shorthand notation
\begin{equation}
\lambda_0 \equiv \frac{1}{4} (g_2^2+ \frac{3}{5}g_1^2)
\end{equation}
for the tree level Higgs quartic coupling, while $\lambda$ is reserved for the quartic coupling containing quantum effects (to be discussed below). Here $g_1$ is the $U(1) \subset SU(5)_{GUT}$ gauge coupling, related to the conventional hypercharge coupling as $g'^2=\frac{3}{5}g_1^2$.

In the present analysis, the running masses of the heavy composites can be taken to be constant;  the running of the Yukawa couplings can be similarly neglected. 
This is a rather good approximation and simplifies many of the formulae.
Solving the whole system of one loop RGEs, we have checked numerically the consistency of this approximation; these effects are taken into account in the computations leading to the results of \S \ref{sec:pheno}.

Clearly, the main changes occur for the TeV scale elementary masses. Our composite spectrum has two characteristic features that may affect the RG evolution considerably:
\begin{itemize}
\item Heavy first and second generation sfermions that enter into the $\beta$-functions of $H_u$ and third generation sfermions via Yukawa couplings, $\beta \propto y_i^2 m_{\t Q_i}^2$. While such Yukawas are parametrically smaller than $y_t$ and $y_b$, these effects are still important because $m_{\t Q_i} \gg m_{\t Q_3}$ ($i=1,2$). Such contributions decrease the masses of the elementary particles.
\item A hybrid Higgs sector with $m_{H_d}^2 \gg m_{H_u}^2$; this induces a one-loop FI term for $U(1)_Y$,
\begin{equation}
 S={\rm Tr}(Ym^2)=m^2_{H_u}-m^2_{H_d}+\sum_i(m^2_{\t q,i}-2 m^2_{\t u,i}+m^2_{\t d,i}-m^2_{\t l,i}+m^2_{\t e,i}) \sim -m_{H_d}^2
\end{equation}
In the RGEs, this contribution increases the masses of fields with positive hypercharge (and vice-versa). The usual sum rule ${\rm Tr} (Ym^2)=0$ is violated and there are strong effects on the third generation sleptons. See also the related discussion in~\cite{Csaki:2008sr}.
\end{itemize}

First, we show in Appendix \ref{app:one-loop-MSSM} that composite fields give  small contributions to the running of $m_{H_u}$ and the squark masses, which is dominated by the third generation elementary fields. The RG running due to the composite fields alone changes $m_{H_u}^2(t)$ marginally, while it contributes to a small decrease of the  squark masses  (e.g. by $\sim 1 \%$ for $\t u_3$). Explicit results at the top scale are shown below.

On the other hand, the running of the third generation slepton masses receives large corrections from the composite fields. This effect is especially important for the soft mass $m^2_{\t l,3}$, whose $\beta$-function can be approximated by
\begin{equation}\label{eq:betaL3}
8 \pi^2\,\frac{dm^2_{\t l,3}}{dt}\approx y_{l,33}^2(m_{H_d}^2+m^2_{\t l,3}+m^2_{\t e,3})-\frac{3}{5}g_1^2M_1^2-3g^2_2M_2^2-\frac{3}{10}g_1^2 S+y_{l,32}^2(m_{H_d}^2+m^2_{\t e,2})\,.
\end{equation}
Now all the loop effects from composites decrease this mass squared. Evaluating it at the scale of the MSSM composites $m_{CW}$, the condition to avoid a tachyonic slepton requires 
\begin{equation}\label{eq:boundstau}
m_{\t l_3}(0) \gtrsim 10^{-2} m_{CW}\,.
\end{equation}
Equivalently, we find a lower bound on the stop mass at the messenger scale,
\begin{equation}\label{eq:boundstop}
m_{\t q,3}(0) \gtrsim \frac{m_{CW}}{10}\,.
\end{equation}
In other words, the requirement of a non-tachyonic elementary slepton implies that the stop cannot be much lighter than $2\,{\rm TeV}$ in our composite model.\footnote{The authors of~\cite{ArkaniHamed:1997ab} obtained a lower bound on the stop from two loop effects of heavy first and second generations; in a general scenario, this restricts the possibility of solving the flavor problem by decoupling. In our composite models such bound is less restrictive than (\ref{eq:boundstop}), which comes primarily from having a composite Higgs. Indeed, in the absence of a composite Higgs, the bound is closer to $m_{\t q,3}(0) \sim m_{CW}/50$. The flavor problem is solved as explained in~\cite{Craig:2009hf}.}

Notice that the strong RG effects on the elementary slepton are a direct consequence of the compositeness of the first SM generations and $H_d$. These parametrically heavy fields decrease $m^2_{\t l,3}$ through both inter-generational mixing and a large $U(1)_Y$ FI term. For the right handed slepton, effects from composite fields are also important but smaller, because now there is a partial cancellation between the contributions from off-diagonal Yukawa couplings and the $U(1)_Y$ FI term. This generically results in an increase of the mass when flowing to the TeV scale.

\subsubsection{Finite effects from composites}

At the scale $m_{CW}$ the MSSM composites are integrated out following \S \ref{subsec:method}. First, since $m_{H_d}^2 \gg m_{H_u}^2$, $H_d$ can be simply integrated out by imposing its equation of motion from Eq.~(\ref{eq:fullVH}),
\begin{equation}
\frac{H_d^0}{H_u^{*0}} \approx \frac{B_\mu}{ m_{H_d}^2}
\end{equation}
valid to lowest order in $H_d/H_u$. In our conventions $B_\mu$ is positive and, in what follows, the phases of the Higgs VEVs are rotated away. Therefore, we see that the ``direction'' of EWSB,
\begin{equation}
\tan \beta  = \frac{v_u}{v_d}\;,\;v_u \equiv \langle H_u^0 \rangle \;,\;v_d \equiv \langle H_d^0 \rangle
\end{equation}
is determined at the scale of the composites to be
\begin{equation}\label{eq:tanbeta2}
\tan \beta = \frac{m_{H_d}^2}{B_\mu} \sim \frac{\lambda_d}{\lambda_u} \gg 1 \,.
\end{equation}
This agrees with the result of Eq.~(\ref{eq:tanbeta1}). Note that, though the $B_\mu$ term generated at the messenger scale induces a mixing between $H_d$ and $H_u$, the fact that $m^2_{H_d}\gg B_\mu$ means that the heavy  Higgs eigenstates are essentially aligned with $H_d$, which can then be integrated out. 

One loop effects from composite fields produce a finite shift to the Higgs mass (analogous to (\ref{eq:toy1})), and leave the quartic coupling unchanged --see Eq.~(\ref{eq:toyCW}).  The  Coleman Weinberg contributions to the effective Higgs potential at the threshold due to the composite fields are
\begin{equation}\label{eq:Veff-stage1}
V_{CW} \supset \left(-\frac{B_\mu^2}{m_{H_d}^2}+ \frac{3}{8 \pi^2} \left[\frac{1}{20} g_1^2 m_{H_d}^2-y_{u,22}^2 m_{\t q,2}^2 \right] \right)|H_u|^2+O(|H|^6)\,,
\end{equation}
where all the objects are evaluated at $m_{CW}$ (roughly, the average mass of the composite fields). The contribution from $\t Q_2$ is subdominant compared to the effect from $H_d$; however, the analogous effect for the stop will be important. Even smaller terms from first generation particles are not shown.  In the theory below the threshold, the coefficient of $|H_u|^2$  in Eq.~\eqref{eq:Veff-stage1} is absorbed into the tree-level parameter $\overline m^2_{H_u}$ at the scale  $m_{CW}$, 
\begin{equation}\label{mHuCorr}
 \overline m^2_{H_u}(m^2_{CW})=\left. m_{H_u}^2-\frac{B_\mu^2}{m_{H_d}^2}+ \frac{3}{8 \pi^2} \left[\frac{1}{20} g_1^2 m_{H_d}^2-y_{u,22}^2 m_{\t q,2}^2 \right]\right|_{Q^2=m_{CW}^2}\,.
\end{equation}
We see that the dominant contribution to the shift of the Higgs mass comes in fact from $B_\mu/m_{H_d}^2$.
The $\beta$-functions for  $\overline m^2_{H_u}$ and the other soft parameters of the theory below the threshold are obtained by  dropping the contributions of the  composites to the MSSM RGEs, while the $\beta$-function for the quartic coupling is modified as in Eq.~(\ref{eq:beta-lambda}).

\subsection{Integrating out the stop and EWSB}\label{subsec:top-scale}

At this stage the effective theory contains $H_u$, third generation sfermions and gauginos, plus the SM matter. The RG evolution and CW corrections are given by the well-known loop effects from third generation particles; gaugino contributions are small. The  dominant contributions of the third generation particles to the CW potential at a scale $Q$ are
\begin{eqnarray}\label{eq:CW3rd}
V_{CW}(Q) &\supset& -\frac{3}{8 \pi^2} (y_{u,33} m_{\t q,3})^2 \left(1-\log \frac{m_{\t q,3}^2}{Q^2} \right) |H_u|^2+\frac{3 }{16\pi^2}y_{u,33}^4 \log\frac{m^2_{\t q,3}}{Q^2}\,|H_u|^4 +\nonumber\\
&-& \frac{3}{16 \pi^2} y_{u,33}^4 |H_u|^4 \left(\log \frac{y_{u,33}^2 |H_u|^2}{Q^2}-\frac{3}{2} \right)\,.
\end{eqnarray}
The first term is the familiar negative shift of the Higgs mass produced by the stop, while the term in the second line comes from loop diagrams with the top quark; A-terms have been neglected. See also~\cite{Carena:1995wu}.

Integrating out the stop/sbottom at $Q=m_{\t q,3}$ produces
\begin{equation}
\overline{m}_{H_u}^2(m_{\t q,3}) = m_{H_u}^2(m_{\t q,3})   -\frac{3}{8 \pi^2} (y_{u,33} m_{\t q,3})^2\,.
\end{equation}
The tree-level parameter $\lambda$ does not receive threshold contributions, though its running changes below $Q=m_{\t q,3}$ due to the  $Q$-dependent logarithm in the quartic term in Eq.~\eqref{eq:CW3rd},
\begin{equation}\label{eq:beta-mod-lambda}
\beta_\lambda \to \beta_{\lambda}-\frac{3 y_t^4}{8\pi^2}\,,
\end{equation}
a known result in the MSSM decoupling limit. After decoupling the squarks of the third generation, there remains the threshold of the gauginos. Their finite one loop corrections are negligible, so that their decoupling is simply implemented by a change in the $\beta$-functions.

Finally we have gathered all the necessary results to study the breaking of $SU(2) \times U(1)$ at the top scale $m_t \approx y_{u,33}\,v_u$. Minimizing the effective potential
\begin{equation}\label{Vtop}
V_{eff} \approx m_{H_u}^2 |H_u|^2+ \frac{\lambda}{2} |H_u|^4- \frac{3}{16 \pi^2} m_t^4 \left( \log \frac{m_t^2}{Q^2}- \frac{3}{2}\right)
\end{equation}
and evaluating at $Q=m_t$ yields
\begin{equation}\label{eq:EW-vacuum}
\lambda(m_t)\, v_u^2 = - m^2_{H_u}(m_t)-\frac{3}{8 \pi^2} y_{u,33}^2 m_t^2\,.
\end{equation}
The resulting expectation value is real due to the absence of phases in the soft parameters. This EWSB condition is key in relating the microscopic parameters to known physics at the top scale. Indeed, following the RG evolution and the threshold corrections of the parameters $m^2_{H_u}$ and $\lambda$ from the messenger scale down to the top scale (as described above), the Higgs VEV $v_u$ is calculated in terms of the high energy parameters, and we have to search for ranges that yield the correct $v_u\sim 174\,{\rm GeV}$. Furthermore, recalling that the Yukawa couplings are fixed in the UV theory (at the scale $M_{\rm{flavor}}$), $\tan \beta$ in Eq.~(\ref{eq:tanbeta2}) is determined in terms of the bottom mass.

Let us summarize how $m_{H_u}^2(m_t)$ and $\lambda(m_t)$ are calculated. The Higgs mass parameter is expressed in terms of the messenger scale parameters by
\begin{equation}
m_{H_u}^2(t) = \sum_i \frac{\partial m_{H_u}^2(t)}{\partial m_i^2(0)}\,m_i^2(0)\,,
\end{equation}
where the soft parameters $m_i^2(0)$ are functions of $(h \hat \mu, x, \lambda_u, h)$ that were described in \S \ref{subsec:param}. Solving Eq.~(\ref{eq:beta-system}) along the different energy ranges and adding up the finite corrections reveals that the dominant contributions are: a) dynamical, from the $B_\mu$ term in Eq.~(\ref{eq:Veff-stage1}); and b) radiative, from stop effects. Other contributions are subleading.

\begin{figure}
\begin{center}

\bigskip
\bigskip
\epsfig{file=  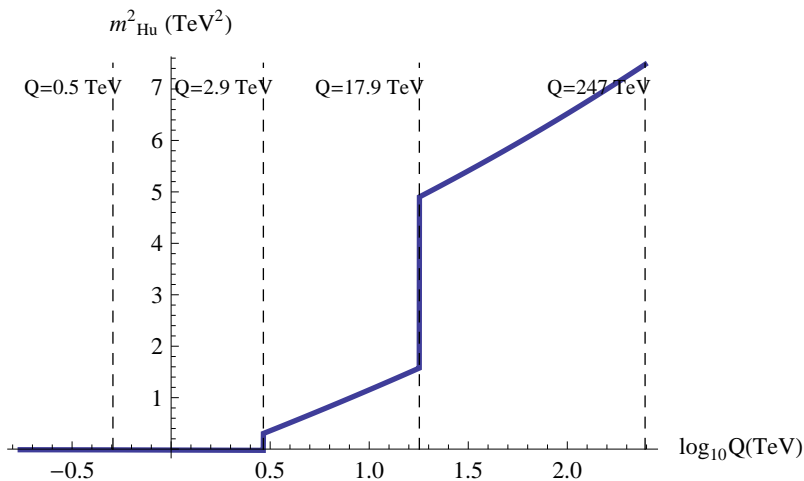 =.8\textwidth}

\caption{RG evolution for $m_{H_u}^2$.} . 
\label{fig:RGPlot}
\end{center}
\end{figure}

More precisely, the Higgs mass in terms of the magnetic superpotential parameters becomes
\begin{eqnarray}\label{eq:analytic-mH}
&&m_{H_u}^2 (t) \approx  \frac{(h \hat \mu)^2}{8 \pi^2} \left( \lambda_u^2 \left(c_{H_u}\,\hat V^{cw}_{H_u H_u^*}+c_{B_\mu} \frac{(\hat V^{cw}_{H_u H_d})^2}{\hat V^{cw}_{H_d H_d^*}}\right)+  \hat V^{cw}_{H_d H_d^*}(h^2 c_{\t q,2}+\lambda_d^2 c_{H_d})+\right.\\
&+&\left.\frac{1}{2} f_{gm}\left[c_{H_u} \sum_a C_a^{H_u} \alpha_a^2+ c_{\t q,3} \sum_a C_a^{q,3} \alpha_a^2+c_{\t u,3} \sum_a C_a^{u,3} \alpha_a^2\right]+ \frac{1}{2} (x f_\lambda)^2 (c_{M_3} \alpha_3^2+c_{M_2} \alpha_2^2)\right)\nonumber\,.
\end{eqnarray}
Here $c_i \equiv \partial m_{H_u}^2(t)/\partial m_i^2(0)$ and all the running objects are evaluated at $t \sim \log \,m_t/(h \hat \mu)$. This makes explicit the dependence on the microscopic parameters. For instance, for 
\begin{equation}\label{eq:good-range}
 h \hat \mu \sim 200\;{\rm TeV}\;,\; \muphi \sim 0.7\;{\rm TeV}\;,\;\t N_c \sim h \sim \lambda_d \sim 1\;,\;\lambda_u \sim 0.1
\end{equation}
the second derivatives are of order
\begin{eqnarray}
c_{H_u} &\sim& 0.78\;,\;c_{B_\mu} \sim -0.85\;,\;c_{\t q, 2} \sim - 2 \times 10^{-3}\;,\;c_{H_d} \sim 3 \times 10^{-3} \nonumber\\
c_{\t q,3} &\sim&-0.16\;,\;c_{\t u,3}\sim -0.12\;,\;c_{M_3} \sim  -0.05\;,\;c_{M_2} \sim 0.1\,.
\end{eqnarray}
In figure \ref{fig:RGPlot} the RG evolution for $m_{H_u}^2$ is shown. The parameters are the same that give  rise to the spectrum in figure \ref{fig:Spectrum1}. Notice that there are three thresholds, where $m_{H_u}^2$ receives finite corrections, with the largest contribution coming from the composite fields  (\ref{mHuCorr}).
 In the first two segments of the evolution from the messenger scale $Q=h\hat \mu$ to the stop scale the slopes are almost equal, which reflects the fact that the running is dominated by the third generation, and the effects of the composites are small. 


With the help of the result (\ref{eq:analytic-mH}), in the next section we will find realistic EWSB vacua in the range (\ref{eq:good-range}), where all the relevant terms contributing to $m_{H_u}^2$ are of the same order of magnitude. As we discuss in Appendix \ref{subsec:tuning}, this will grant that EWSB occurs quite naturally.

The quartic coupling is calculated starting from the tree level coupling $\lambda_0$ and solving the modified $\beta$-functions (as in Eq.~(\ref{eq:beta-mod-lambda})):
\begin{equation}
\lambda(m_t) \approx \lambda_0- \frac{3}{8 \pi^2} y_{u,33}^4\,\log\frac{m_t^2}{m_{\t q,3}^2}\,.
\end{equation}
Contributions from the MSSM composites are negligible. Using this, the one loop mass of the light physical Higgs (predominantly ${\rm Re} (H_u^0)$) is
\begin{equation}
m_h(m_t)^2 \sim 2 \lambda(m_t) v^2\,,
\end{equation}
(also, see e.g.~\cite{Okada:1990vk,Ellis:1990nz,Haber:1990aw}). Our model has a stop around $3 \,{\rm TeV}$, yielding a neutral Higgs around $140\,{\rm GeV}$. Higher loop effects (not taken into account here) tend to decrease this value. The full spectrum and low energy predictions are discussed next.

\section{Low energy phenomenology}\label{sec:pheno}

In \S \ref{sec:eft} we have already outlined general properties of the low-energy physics that we expect in our model. This section presents results for the spectrum from direct numerical calculations, solving the coupled RGEs in the step-wise procedure, and including finite effects, running of Yukawas and decoupling of heavy particles at various thresholds. We stress that these are one loop results (certain higher loop effects are also included, when they can be resummed into one loop contributions in the effective potential). A more detailed two loop analysis is postponed to a future work.

We also perform a scan of the allowed parameter space under some simplifying assumptions explained below. The NLSP (at the one loop level) is found in different energy ranges, and the results are presented in the $M_1 - \mu$ plane, figure \ref{fig:M1muTanb10}, which may be used for comparison with projected exclusion plots from Tevatron and (future) LHC searches~\cite{Meade:2009qv}.

\subsection{Characteristics of the spectrum}\label{subsec:signatures}

Let us begin by summarizing the main features of the spectrum:
\begin{itemize}
\item Composites (i.e. first and second generation sfermions and $H_d$) have masses 
\begin{equation}
m_{\t Q_i}^2 \sim m_{H_d}^2 \sim m_{CW}^2 \sim \frac{h^2\t N_c}{16 \pi^2} (h \hat \mu)^2 \,.
\end{equation}
\item Elementaries (i.e. third generation sfermions and $H_u$) have masses generated from standard gauge-mediation at two-loops,  
$$
\t N_c\left(\frac{g^2}{16 \pi^2}\right)^2 (h \hat \mu)^2\,.
$$
\item Gauginos and higgsinos have masses proportional to $\mu_\phi$. 
\end{itemize}

\begin{figure}
\begin{center}

\epsfig{file=  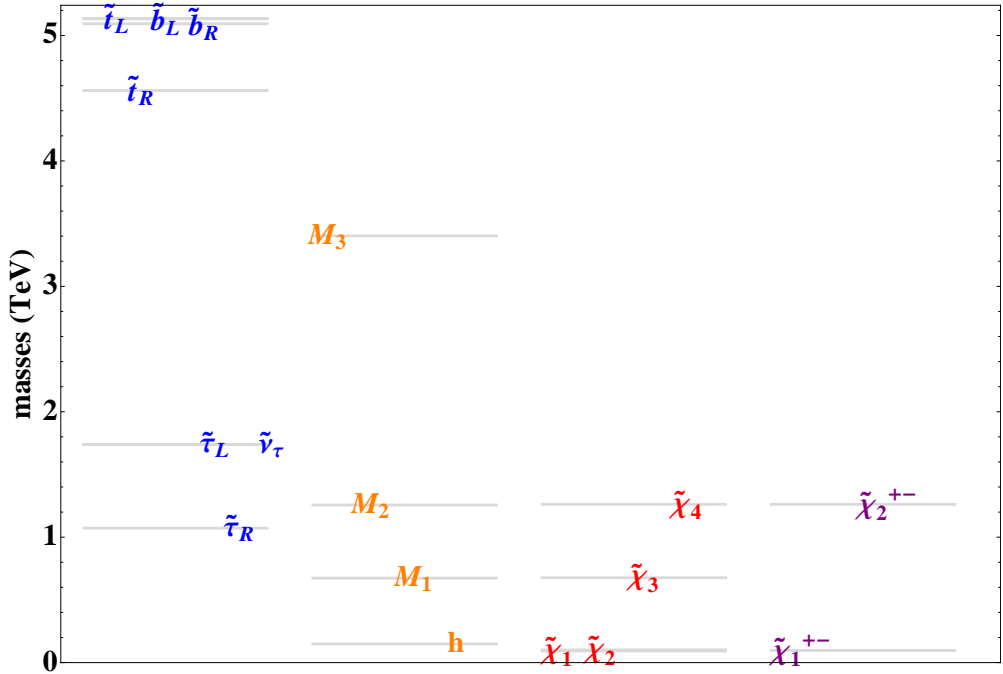 ,width=.8\textwidth}
\bigskip
\bigskip
\bigskip

\epsfig{file=  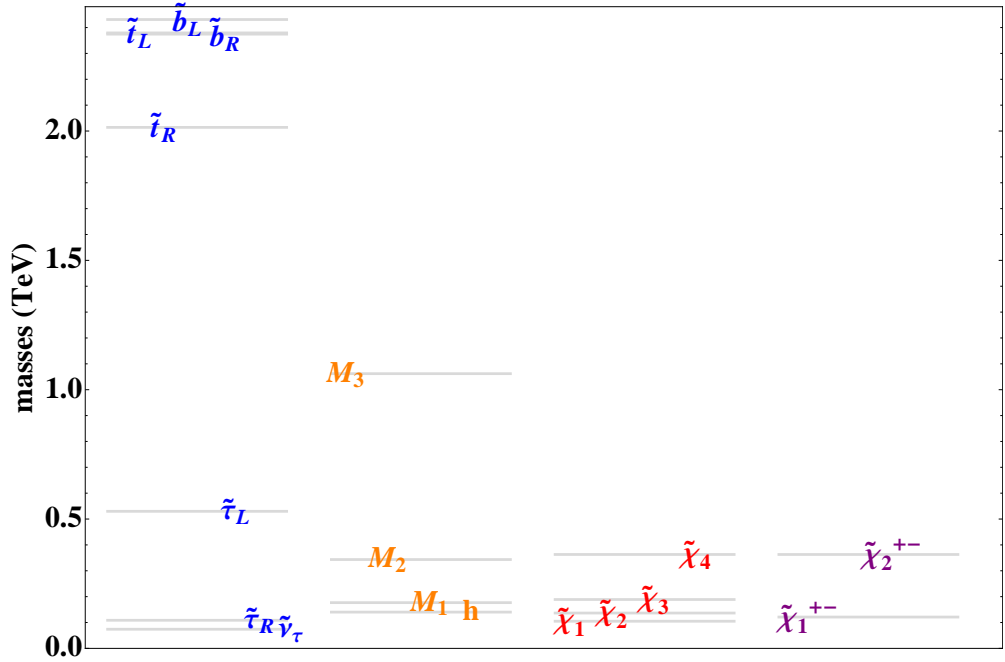 ,width=.8\textwidth}
\caption{Sample spectra focusing on the masses around $1$TeV, for $\t N_c=3,$ tan$\beta\sim 7$ with higgsino NLSP and $\t N_c=1$, tan$\beta\sim 6.5$ with sneutrino NLSP.} . 
\label{fig:ZoomedSpectrum}
\end{center}
\end{figure}

First, there is an inverted hierarchy, with the first two generation squarks and sleptons being much heavier than their counterparts from the third generation. For parameter ranges explored here, these masses are of order $ 10 - 20 $ TeV. Also, from the hierarchy $m_{H_d}^2 \gg m_{H_u}^2$, the Higgs scalar eigenstates $(A^0,\,H^{\pm}, H^0)$ are much heavier than $h^0$ and have masses of the same order as the other composites.  Ignoring such heavy particles, two detailed sample spectra are shown in figure \ref{fig:ZoomedSpectrum}, for $\tan \beta = 6.5-7$ in two parameter regimes, where the NLSP is neutralino and sneutrino, respectively\footnote{The complete spectrum for the higgsino NLSP case is shown in figure \ref{fig:Spectrum1}.}. 
The parameter choices for these cases are $\t N_c=3$ and
\begin{equation}
\begin{aligned}
\t\chi_1^0 \hbox{ NLSP}: & \quad \lambda_d \approx h \approx .65 \,,\quad 
 \hat \mu \approx 471 \,\hbox{TeV}\,,\quad 
\mu_\phi \approx  2.6 \, \hbox{TeV}\,,\quad 
\lambda_u \approx 0.14\cr
\t\nu \hbox{ NLSP}: & \quad \lambda_d \approx h \approx 1.4\,,\quad 
 \hat \mu \approx \ 91 \,\hbox{TeV}\,,\quad 
\mu_\phi \approx 1.0 \, \hbox{TeV}\,,\quad 
\lambda_u \approx 0.29 \,.
\end{aligned}
\end{equation}
Higher loop effects could give important corrections, particularly in the case of the sneutrino NLSP (see discussion below). It should be noted that the sneutrino NLSP case above could be unstable towards decay into other vacua (see \S\ref{subsec:newvacua}); however, it is possible that $\mu_\phi$ effects which were ignored in the tunneling computations could enhanced the lifetime of the ISS vacuum. Also, for higher values of tan$\beta$ the sneutrino region moves to lower values of $h$ guaranteeing a suppressed decay rate of the vacuum.


Let us now focus on the fermion masses. It is possible to choose the microscopic parameters such that both gaugino and higgsino masses are real and positive, so there are no CP violating effects. (We remind the reader that we work in the usual convention where $\langle H_u^0 \rangle$, $\langle H_d^0 \rangle$ and $B_\mu$ are real and positive.)

The suppression factor $\lambda_u$, coming from a dimension 5 operator in the electric theory, implies that in general the NLSP is mostly higgsino. There are however, certain parameter regimes where the sneutrino becomes the NLSP, in agreement with the analysis of Eq.~(\ref{eq:betaL3}), (\ref{eq:boundstau}). In these cases there is also a slightly heavier stau. Some comments about fine-tuning in this setup can be found in Appendix \ref{subsec:tuning}.


\subsection{Parameter space and NLSP}\label{subsec:ParameterPlots}

In order to understand the range of predictions of the model,
it is important to systematically scan the parameter space. A full analysis of the parameter ranges is left for the future; here we restrict to the case $h \approx h_i \approx \lambda_d$. Then the input parameters at the messenger scale are
\begin{equation}
\lambda_d\,,\  h\hat \mu\,,\  \lambda_u\,,\ \t N_c\ \hbox{ and } \mu_\phi \,. 
\end{equation}

The first consistency condition is the absence of tachyons; as explained around Eq.~(\ref{eq:boundstau}), these could come primarily from light third generation fields. This implies that $h$ cannot be much larger than one. Next, predicting the correct EWSB vacuum can be used to fix the messenger mass $h \hat \mu$ in terms of the other parameters. This relation follows from Eqs.~(\ref{eq:EW-vacuum}) and (\ref{eq:analytic-mH}). 

Furthermore, the top and bottom mass ratio fixes a relation between $\tan\beta$ and the Yukawa couplings. In scanning the parameter space it is useful to keep the Yukawas, and thus  $\tan\beta$ fixed.  
Given the approximate relation Eq.~(\ref{eq:tanbeta2}),  $\lambda_d$ can be eliminated in favour of  $\lambda_u$. This motivates scanning the parameter space in terms of $\lambda_u$ and $\muphi$ alone, for  a given value of $\tan\beta$. The light masses which are most sensitive to such parameters are the higgsino and bino masses, respectively. So we present our results in the $\mu$-$M_1$ plane.

\begin{figure}
\begin{center}
\epsfig{file=  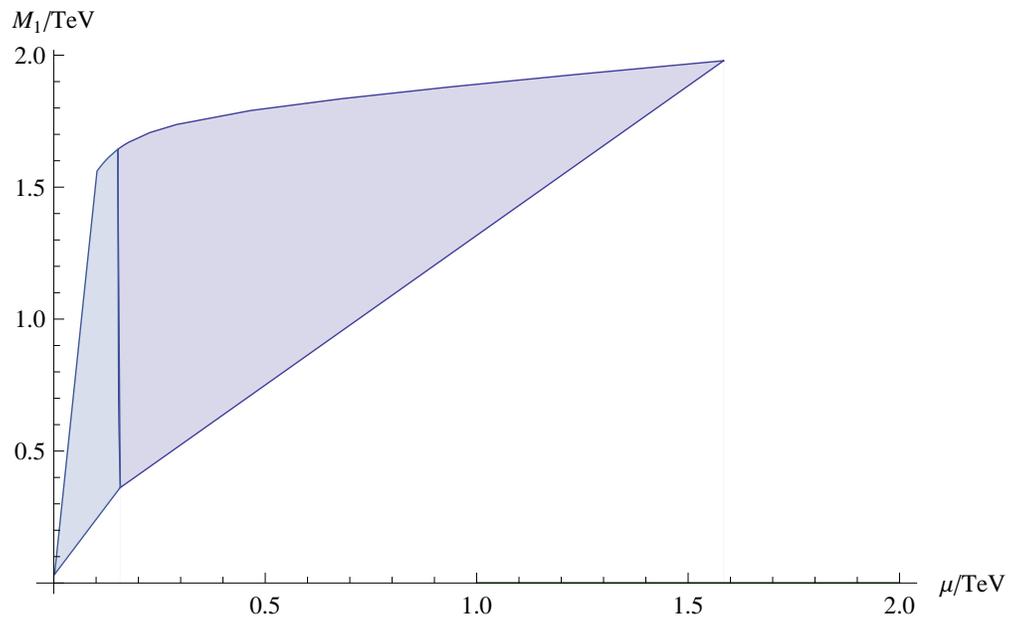,width=.8\textwidth}

\caption{$\mu$ versus $M_1$ parameter space for $\t N_c=3$ and tan$\beta\sim 6.5$. The region consistent with EWSB, absence of tachyons and  higgsinos heavier than $150$ GeV is depicted in blue. The teal region indicates all models that have higgsinos with a mass below that bound -- which is only orientative. For $\t N_c=3$ there is no region with sneutrino NLSP. Two loop effects have been neglected in this analysis.}
\label{fig:M1muTanb10}
\end{center}
\end{figure}

The parameter scan in figure \ref{fig:M1muTanb10} for $\t N_c=3,$ $\tan\beta\sim6.5$ shows four regions in the $\mu$-$M_1$-plane: the purple region is excluded as there are tachyonic sleptons. The boundary to the right comes from requiring $\lambda_u \lesssim 0.4$. Extending the scan further to the right is possible, but such larger values would not be generated naturally from the UV electric theory. The blue region corresponds to models that have consistent EWSB, no tachyons and an NLSP which is predominantly higgsino-like.  We have distinguished the teal region out of the blue region; this corresponds to models that are consistent with EWSB and that have a neutralino lighter than $150$ GeV. Moving closer to the origin, the bino component of the NLSP becomes larger. These models could be ruled out by Tevatron~\cite{Meade:2009qv}. Finally, the thin green region corresponds to consistent models with sneutrino NLSP.


We see that in most cases the NLSP is mostly higgsino-like, which can be understood from $\mu < M_1, M_2$. Writing the lightest neutralino as a linear combination 
\begin{equation}
\t\chi_1^0 = \sum_{i=1}^4 N_{1, i} \t\psi_i^0 \;,{\rm where}\;\psi_i^0 = (\t B, \t W, \psi_{H_u}, \psi_{H_d})\,.
\end{equation}
We have $|N_{13}|, |N_{14}|> |N_{11}|, |N_{12}|$, and the sign of the fourth eigenvalue is sign$(N_{14}) <0$.~\cite{Meade:2009qv} then suggests that the higgsino-like NLSP should decay mostly to $Z$'s. The detailed structure of decays and final states depends also on the mass splitting between the lightest neutralino and chargino. 

On the other hand, the case of sneutrino NLSP is also of particular
interest, as this is not easy to realize in perturbative scenarios.
For the region of parameters that we studied, and computing the masses
using tree level (soft plus Higgs induced mass terms) and one loop RG effects, whenever a slepton was the NLSP, it was the
sneutrino. This intriguing possibility has been recently studied
in~\cite{Katz:2009qx}. This might seem surprising at first sight, because the gauge mediated slepton mass $m_{\tilde l_3}^2$ (which determines the sneutrino mass eigenvalue) is larger than $m_{\tilde e_3}^2$. However, in our case we
have strong RG effects from the composite generations, coming from
off-diagonal Yukawas and a large $U(1)_Y$ FI term, caused by the
composite $H_d$ --see Eqs.~(\ref{eq:betaL3}) and (\ref{eq:boundstau}).
This produces an inversion in the hierarchy between $m_{\tilde l_3}^2$
and $m_{\tilde e_3}^2$. The sneutrino and the lightest stau mass
eigenvalues become similar, the latter turning a bit heavier due to
the leptonic Yukawa coupling and the Higgs D-term contributions. 

\begin{figure}
\begin{center}
\epsfig{file=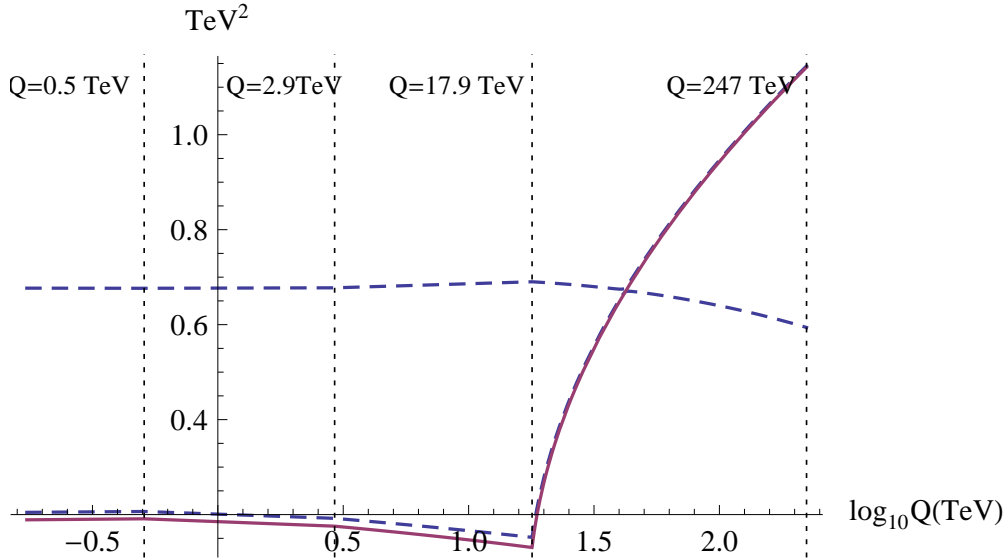,width=.8\textwidth}
\caption{RG evolution for the stau mass eigenvalues (dashed lines) and
sneutrino mass (continuous line).}
\label{fig:RGPlot2}
\end{center}
\end{figure}

Since
the splitting between these masses ends up being quite small (around
15 GeV), two loop corrections might become important and could alter
the picture; we leave a more detailed analysis for future work. The
running mass eigenvalues --neglecting the running of the Higgs VEVs--
for the same choice of parameters that led to
Figs.~\ref{fig:Spectrum1}~and~\ref{fig:RGPlot}   are shown in
Fig.~\ref{fig:RGPlot2}. 

\subsection{Concluding remarks}

Let us conclude with some brief remarks about our construction and possible future directions. We have argued that SQCD with flavors and an adjoint, plus an appropriate superpotential, can simultaneously generate dynamically the electroweak scale, explain the flavor hierarchies (as in~\cite{Franco:2009wf,Craig:2009hf}) and produce a realistic low energy spectrum. It is rather intriguing that the complicated structure of the MSSM can originate from a quite simple microscopic theory. 

It is necessary to point out that, since the electric theory is vector-like, it suffers from the existence of extra matter near the compositeness scale. We have not addressed this aspect here, because it does not affect the EW effective theory or the soft parameters. It would be useful to understand better their role, and to find models where unification occurs naturally. 

The theory at the TeV scale also exhibits an interesting phenomenology, which we have begun to explore in this work. A striking feature is that a small number of microscopic couplings controls all the soft parameters and EW scale; this produces nontrivial relations between the MSSM sectors, that can also vary along the parameter space of the model. We have restricted to the case where the cubic couplings produced by Seiberg duality are all of the same order of magnitude, $h_i \sim \lambda_d$. New effects are expected away from this subspace. A more detailed analysis, including possible collider signatures, is left to future work.

\subsection*{Acknowledgements}

We would like to thank 
N.~Craig, 
C.~Csaki,
M.~Dolan,
R.~Essig,
S.~Franco,
K.~Intriligator,
S.~Kachru, 
J.~Marsano,
M.~Peskin,
E.~Silverstein,
M.~Strassler
and J.~Wacker
for very interesting discussions and comments. In particular, we thank R.~Essig for discussions regarding two loop effects and direct search bounds. S.S.-N., C.T. and G.T. are supported in part by NSF grant PHY-05-51164 at the KITP.
G.T. is supported by the US DOE under contract number DE-AC02-76SF00515 at SLAC.
C.T. is supported by  in part by MICINN through grant
FPA2008-04906 and by both MICINN and the Fulbright Program through grant 2008-0800.
S.S.-N. thanks the Caltech theory group for their generous hospitality. G.T. and C.T. thank the KITP for hospitality.

\appendix 

\section{One loop computations at the messenger scale}\label{app:one-loop}

The analysis of the one loop effects in the ISS model with quadratic deformation has already been performed in~\cite{Essig:2008kz}, and the first part of this appendix summarizes these results (with the addition of background Higgs fields). The second part presents details of the soft term, and the calculation of the one loop $\mu$-term.

\subsection{Single-sector Coleman-Weinberg contributions}

We perform, as in~\cite{Intriligator:2006dd}, a calculation of the contributions to the one-loop Coleman-Weinberg potential of the heavy messenger fields $\rho,\tilde\rho,Z,\tilde Z$, which allows us to understand the stabilization of $X$  and the computation of soft terms for the Higgs fields and composite generations. Namely, $X$, the Higgs fields and the composite MSSM generations are treated as background fields, and the heavy messengers are integrated out. 

From Eqs.~(\ref{eq:Mf}) and (\ref{eq:Wfullmag}), the relevant superpotential terms are
\begin{equation}\label{eq:Wiss2}
W_{mag}={\rm tr}\;\left( \begin{matrix} \rho& Z\end{matrix} \right) \left( \begin{matrix}  h X+h_i \Phi_i^{SM}  + \lambda_d H_d & h\hat \mu \\ h \hat \mu & h^2 \muphi+ \lambda_u H_u\end{matrix} \right) \left( \begin{matrix} \tilde \rho\\ \tilde Z\end{matrix} \right)\,,
\end{equation}
giving a field-dependent supersymmetric mass matrix
\begin{equation}\label{eq:appMf}
M_f=\left( \begin{matrix}   h X+h_i \Phi_i^{SM}  + \lambda_d H_d 
& h\hat \mu \\ h \hat \mu 
& h^2 \muphi+ \lambda_u H_u\end{matrix} \right) \,.
\end{equation}
To simplify our formulae, we absorb the composite generations into $X$ (i.e. $h X+h_i \Phi_i^{SM} \to hX$) since their dependence can be easily restored at the end.

The only background superfield with an $F$-term is $\tr X$. Therefore the bosonic mass matrix for $(\rho, Z)$ is
\begin{equation}
M_b^2=\left(
\begin{array}{cc}
M_f^\dagger M_f & -h^*F_X^*\\
-hF_X & M_f M_f^\dagger
\end{array}
\right) \,,
\end{equation}
where
\begin{equation}
-F_X^*=h\left(
\begin{array}{cc}
-\hat \mu^2+h \muphi\,X & 0\\
0 & 0
\end{array}
\right).
\end{equation}

Regrouping the fields as
\begin{equation}
\hat{\psi}=\left(\;\psi_{\rho}\;\;\;\psi_{Z}\;\right)^T\;\;\;\;\;\; \hat{\tilde{\psi}}=\left(\;\psi_{\tilde{\rho}}\;\;\;\psi_{\tilde{Z}}\;\right)^T \;\;\;\;\;\;\hat{\phi}=\left(\;\rho\;\;\;Z\;\;\;\tilde{\rho}^*\;\;\;\tilde{Z}^*\;\right)^T
\end{equation}
the mass terms give
\begin{equation}
\mathcal{L}_{mass}=-\hat{\tilde{\psi}}M_f\hat{\psi}-h.c.-\hat{\phi}^\dagger M_b^2\hat{\phi}\,.
\end{equation}
The messenger mass matrices can be diagonalized by unitary matrices $U_f$, $\tilde{U}_f$ and $U_b$ such that
\begin{equation}\label{eq:Umatrices}
\psi=U_f\hat{\psi}\;\;\;\;\;\;\tilde{\psi}=\tilde{U}_f\hat{\tilde{\psi}}\;\;\;\;\;\;\phi=U_b\hat{\phi}
\end{equation}
where $\psi$, $\tilde{\psi}$ and $\phi$ are messenger mass eigenstates.  The quadratic Lagrangian for the messengers is therefore of the canonical form
\begin{eqnarray}
\mathcal{L}_{mess}&=&-\sum_{a=1}^4\phi_a^\dagger\left(D^2+\tilde{m}_a^2\right)\phi_a+\nonumber\\
&+&\sum_{a=1}^2\left(\bar{\psi}_ai\bar{\sigma}^\mu D_\mu\psi_a+\bar{\tilde{\psi}}_ai\bar{\sigma}^\mu D_\mu\tilde{\psi}_a-m_a(\tilde{\psi}_a\psi_a+\bar{\tilde{\psi}}_a\bar{\psi}_a)\right).
\end{eqnarray}

The fermionic and bosonic mass eigenvalues are
\begin{eqnarray}\label{nonSUSYmasses}
m^2&&= |h\hat \mu|^2+\frac{1}{2}|hX+\lambda_d H_d|^2+\frac{1}{2}|h^2 \muphi+ \lambda_u H_u|^2\\ 
&&+\frac{1}{2}\sigma\sqrt{\left(|hX+\lambda_d H_d|^2-|h^2 \muphi+ \lambda_u H_u|^2\right)^2+4|h\hat \mu(hX+\lambda_d H_d)^*+(h\hat \mu)^*(h^2 \muphi+ \lambda_u H_u)|^2}\nonumber\\
\tilde{m}^2&& = |h\hat \mu|^2+\frac{1}{2}|hX+\lambda_d H_d|^2+\frac{1}{2}|h^2 \muphi+ \lambda_u H_u|^2+\frac{1}{2}\eta|(h\hat \mu)^2-h^2 \muphi (hX+\lambda_d H_d)|\nonumber \\
 && +\frac{1}{2}\sigma\Big[\left(|hX+\lambda_d H_d|^2-|h^2 \muphi+ \lambda_u H_u|^2+\eta|(h\hat \mu)^2-h^2 \muphi (hX+\lambda_d H_d)|\right)^2+\nonumber\\
&& 4|(h\hat \mu)(hX+\lambda_d H_d)^*+(h\hat \mu)^*(h^2 \muphi+ \lambda_u H_u)|^2\Big]^{1/2}\,.
\end{eqnarray}
Here $\sigma = \pm$, $\eta =\pm$; the fermion masses have multiplicity $4N_c\tilde{N}_c$ and the complex bosons have multiplicity $2N_c\tilde{N}_c$. In the main part of the paper we set $\t N_c=1$.

One loop effects from integrating out the messengers (at their average mass $h\hat \mu$) give the effective potential
\begin{equation}\label{eq:softCW}
V_{CW}=\frac{1}{32\pi^2}\left(\sum_{i=1}^4 \t m_i^4 \left(\log \frac{\t m_i^2}{(h \hat \mu)^2}-\frac{3}{2}\right)-2\sum_{a=1}^2 m_a^4 \left(\log \frac{ m_a^2}{(h \hat \mu)^2}-\frac{3}{2}  \right)\right)\,.
\end{equation}

\subsection{Bosonic soft terms at the messenger scale}\label{appsubsec:soft}

Soft terms are identified by expanding the potential around $\langle h X \rangle$ (Eq.~(\ref{eq:Xvev})), while the VEVs for the Higgs fields are negligible at this stage. For the bosonic fields this gives
\begin{eqnarray}\label{eq:expansionVCW}
V_{CW}(h\hat \mu)&=&m_{\t Q_i}^2\,|\t Q_i|^2+m_{H_u}^2 |H_u|^2+ m_{H_d}^2|H_d|^2-
(B_\mu  H_u H_d+c.c.)\\
&-& A_u^{ij}\,(\t Q_i H_u \t Q_j)- A_d^{ij}\,(\t Q_i H_d \t Q_j)
-A_u^{'\,ij}\,(\t Q_i H_d^\dag \t Q_j)-A_d^{'\,ij}\,(\t Q_i H_u^\dag \t Q_j)-c.c.+ \ldots,\nonumber
\end{eqnarray}
where $\t Q_i$ denote the first two composite MSSM generations, arising form the lowest components of the superfields $\Phi_i^{SM}$ in (\ref{eq:appMf}). Also, `$\ldots$' are quartic and higher order corrections, which are much smaller than the tree level D-term potential for $H$.

The soft parameters depend on the $X$ VEV. For $\muphi=0$, the pseudo-modulus is stabilized at the origin and the soft terms were given in \S \ref{subsec:dsb}; the A-terms vanish in this limit. However, as explained in \S \ref{subsec:mu}, in this case there is an unbroken R-symmetry that forbids gaugino and higgsino masses. Switching on the quartic deformation $\muphi \neq 0$ shifts the metastable vacuum to 
\begin{equation}
\langle h X \rangle \approx \frac{8 \pi^2}{\t N_c(\log 4-1)} \mu_{\phi}^* \,.
\end{equation}
The R-symmetry is both spontaneously and explicitly broken, the first one dominating.

It is convenient to introduce the dimensionless R-symmetry breaking order parameter
\begin{equation}
x \equiv \left|\frac{\langle X \rangle}{\hat \mu}\right|\,.
\end{equation}
The mass terms at the messenger scale receive the following corrections to lowest order in $x$,
\begin{eqnarray}\label{eq:SoftMuPhi1}
m_{H_d}^2 & =&  \frac{\lambda_d^2 \t N_c}{ 8 \pi^2}\left( (\log 4 -1) 
- \frac{1}{3} (12 \log 2 -7) x^2 
	+\mc O(x^4)\right)(h \hat \mu)^2\nonumber\\
m_{H_u}^2  &=& \frac{\lambda_u^2\t N_c}{8 \pi^2}\,\left(-  (1-\log 2 ) 
	+   \frac{5}{6} (4-3\log 2) x^2 + \mc O(x^4) \right) (h \hat \mu)^2   \nonumber\\
B_\mu &=&\frac{\lambda_u \lambda_d \t N_c}{ 8 \pi^2} \left(  (1-\log 2)
- \frac{2}{3} (5-6 \log 2 ) x^2 
	+\mc O(x^4)\right)(h \hat \mu)^2 \,.
\end{eqnarray}
CW masses for the composite generations are obtained from $m_{H_d}^2$ by the replacement $\lambda_d \to h_i$. For simplicity, in this paper we have focused on the case in which all the trilinear couplings generated by Seiberg duality are of the same order, namely, $\lambda_d \sim h_i \sim h$.

The mass squared for $H_u$ starts tachyonic for small $x$, but then becomes positive for $x \gtrsim 0.5$. In fact, this is the regime where realistic gaugino and higgsino masses are generated --these are explained below. For such values of $x$, the nonlinearities from (\ref{eq:softCW}) become important, and higher order terms have to be added to (\ref{eq:SoftMuPhi1}). 

It turns out to be useful to define a dimensionless potential from the one-loop CW potential (second term in Eq.~(\ref{eq:Veff})), as follows:
\begin{equation}\label{eq:defVhat}
\frac{(h \hat \mu)^4}{8 \pi^2} \hat V^{cw}(h,x)\;\;\equiv\frac{1}{64 \pi^2}\,{\rm Str} \;\mc M^4\,\left(\log \frac{\mc M^2}{(h \hat \mu)^2}-\frac{3}{2}\right) \Big|_{\lambda_d =\lambda_u=h_i=1}.
\end{equation}
Here $\mc M$ is the (field-dependent) mass matrix of the messengers. The trilinear couplings $\lambda_d,\lambda_u,h_i$ are set to one because in future formulae we will indicate the explicit dependence of $\hat V^{cw}$ --and hence of the soft parameters-- on them; this  dependence will be given simply by loop counting factors proportional to $\lambda_d,\lambda_u,h_i$. 

Analogously, the gauge-mediated two loop potential (third term in Eq.~(\ref{eq:Veff})) can be  written as
\begin{equation}
V_{GM}^{2-loop}= f_{gm}(h,x)\,(h \hat \mu)^2 \sum_{i,\,a}\,C_a^{i} \left(\frac{g_a^2}{16 \pi^2}\right)^2\,|\phi_i|^2
\end{equation}
for the light scalars $\phi_i$, where $C_a^{i}$ denotes the quadratic Casimir for the group labelled by $a$  in the corresponding representation of $\phi_i$. In the limit in which R-symmetry is restored, $f_{gm}(h,0) \sim 1$ (more precisely, it is proportional to the number of messengers); then it receives small corrections from nonzero $x$.

The soft masses are computed from the second derivatives of $\hat V^{cw}$,
\begin{eqnarray}\label{eq:SoftMuPhi2}
m_{H_d}^2 & =&  \frac{\lambda_d^2 }{ 8 \pi^2}\, \hat V^{cw}_{H_d H_d^*}(h,x)\,(h \hat \mu)^2\nonumber\\
m_{H_u}^2  &=& \frac{\lambda_u^2}{8 \pi^2}\,\hat V^{cw}_{H_u H_u^*}(h,x)\, (h \hat \mu)^2   \nonumber\\
B_\mu &=&- \frac{\lambda_u \lambda_d }{ 8 \pi^2}\, 
	\hat V^{cw}_{H_u H_d}(h,x)\,(h \hat \mu)^2 \,.
\end{eqnarray}
Two loop gauge mediated effects are important in $m_{H_u}^2$. These were parametrized in Eq.~(\ref{eq:mHu-param}) in terms of the dimensionless function $f_{gm}(h,x)$, computed with the help of~\cite{private}. $f_{gm}$
is of order of the number of $\bf{5}+\bf{\overline 5}$ messenger pairs, and here corrections from $x$ are not very important.

Regarding the A-terms, if they are consistent with the tensor product of $SU(5)_{SM}$ representations, their analytic expressions at small $x$ are
\begin{eqnarray}\label{eq:A-terms}
A_u&=&\frac{h^2\t N_c\lambda_u}{24 \pi^2}(6\log 2 -5) x\,
h\hat \mu +\mc O(x^2)\nonumber\\
A_d&=&\frac{h^2\t N_c\lambda_d}{120 \pi^2}(41-60\log 2)x^3\,
 h\hat \mu +\mc O(x^4) \nonumber\\
A_u'&=&\frac{h^2\t N_c \lambda_d}{48 \pi^2}(12\log 2 -7)x\,
h\hat \mu+\mc O(x^2)\,\nonumber\\
\nonumber\\
A_d'&=&\frac{3h^2\t N_c\lambda_u}{160 \pi^2}(43-60\log 2)x^3\,
h\hat \mu+\mc O(x^4)\, .
\end{eqnarray}
In the example considered in this work, where the messenger fields transform in $SU(5)$ representations ${\bf 5}$ and ${\bf \bar{5}}$, only the $A_d$ and $A_d'$ trilinear couplings will be nonzero.

\subsection{Higgsino and gaugino masses}\label{subsec:appmu}

In this subsection, we briefly discuss the explicit one-loop effects that generate gaugino and higgsino masses.

The $\mu$ term is computed from a one-loop diagram with external fermion legs $H_u$ and $H_d$, and with $\rho$ and $Z$ running in the loop. There is a factor of $\lambda_u \lambda_d$ from the vertices, and also a factor of $h \hat \mu$ coming from the loop. 


Following the notation of the Appendix in~\cite{Essig:2008kz}, the Feynman diagram gives
\begin{equation}\label{eq:mu2}
\mu=-\frac{\lambda_u \lambda_d}{16 \pi^2}\,\sum_{i=1}^2\,\sum_{j=1}^4\,I(\t m_j, m_i) \left(U_{b,j4} U^*_{b,j1} U^*_{f,i2} \t U^*_{f,i1}+U_{b,j3} U^*_{b,j2}  U^*_{f,i1} \t U^*_{f,i2} \right)\,,
\end{equation}
where
\begin{equation}\label{eq:I-def}
I(\tilde m_j,m_k)=m_k \left[\ln\left(\frac{\Lambda_{cutoff}^2}{m_k^2}\right)-\frac{\tilde{m}_j^2}{\tilde{m}_j^2-m_k^2}\ln\left(\frac{\tilde{m}_j^2}{m_k^2}\right)\right].
\end{equation}
The one loop $\mu$ term is finite, independent of the cutoff scale $\Lambda_{cutoff}$. Evaluating this gives
\begin{equation}\label{eq:mu3}
\mu \approx \lambda_u \lambda_d\, \t N_c\,\frac{\langle h X \rangle}{16 \pi^2} \approx \lambda_u \lambda_d\,\muphi\,,
\end{equation}
which vanishes in the limit in which the $U(1)_R$ is unbroken. This can be understood from the R-charge assignments of Eq.~(\ref{eq:R-assign}).

\vskip 3mm

Gaugino masses are computed again from one-loop diagrams with scalar and fermion messenger fields $\rho,Z$ running the loop. Quoting \cite{Essig:2008kz}, for a strongly coupled sector with a trivial magnetic group,
\begin{equation}\label{eq:mg}
m_{\lambda}=\,-\frac{2g^2}{16\pi^2}\,\sum_{c=1}^2\sum_{d=1}^2\sum_{j=1}^4\sum_{k=1}^2\, (U_f^*)_{kc}\,(\tilde U_f^*)_{k,d}\,(U_b)_{jc}\,(U_b^*)_{j,d+2}\,I[\tilde m_j,m_k]\sim g^2\mu_\phi.
\end{equation}
As expected, they are proportional to the R-symmetry breaking parameter $\mu_\phi$.

\section{Quantum effects from MSSM fields}\label{app:one-loop-MSSM}

In this appendix we present some formulae used in the computation of quantum effects below the messenger scale, produced by loops of MSSM fields. This includes corrections calculated from Eq.~\eqref{eq:VCWgen}, as well as from the RG flow. We also estimate the size of the two-loop effects on the light soft masses.

The fields that dominate the Higgs-dependent contributions to the effective potential are, from Eq.~\eqref{eq:VCWgen}, those with larger soft masses and/or stronger couplings to the Higgs in their mass matrices.  Since the first two generations have extremely heavy soft masses, their contributions to $V_{CW}$ are important (actually, the second generation dominates). 

Analogously, the heavy neutral and charged Higgs mass eigenvalues have to be taken into account. In the third generation, it suffices to consider the squark, sbottom and top quark. Effects from $A$-terms are subdominant: for the composite generations they are much smaller than the other soft masses, while for the elementary third generation they are generated through gauge mediation and hence negligible. Finite effects from off-diagonal Yukawa couplings are also very small, and will not be included; they will be taken into account in the running of the couplings.

For instance, the Higgs-dependent mass of $\t q_L$ in the generation $i$ reads
$$
m^2_{\t q,i,L} =m^2_{\t q,i}+y^2_{u,ii}|H^0_u|^2+\frac{1}{4}\Big(g_2^2-\frac{1}{5}g_1^2\Big)\Big(|H^0_d|^2-|H^0_u|^2\Big)\,,
$$
where $m^2_{\t q,i}$ is the soft mass and $y_{u,ii}$ is the corresponding diagonal element in the Yukawa matrix. The $H_d$-dependent term in this mass introduces quantum corrections to the down-type Higgs parameters. However, since this composite Higgs has  quite a heavy mass of the order of $m_{CW}$, it can in practice be neglected. On the other hand, the $H_u$-dependent mass term gives large contributions, as we have discussed in \S \ref{sec:eft} and \S \ref{sec:pheno}. The other sparticle masses may be found in e.g.~\cite{Baer:2006rs}.

\subsection{One loop MSSM $\beta$-functions}

For completeness, we reproduce below the one loop MSSM REGs for the soft parameters, including inter-generational mixing, but ignoring A-terms. These are taken from~\cite{Martin:1993zk}; here $t=\log(Q/h\hat \mu)$, $Q$ being the renormalization scale:
\begin{align*}
\frac{d\mu}{dt}=&\frac{\mu}{8\pi^2}\Big(-\frac{3}{5}g_1^2-3g_2^2+\tr(y_{l}^2+3y_d^2+3y_u^2)\Big),\\
\frac{dB_\mu}{dt}=&\frac{\mu^2}{16\pi^2}\Big(-\frac{3}{5}g_1^2M_1-3g_2^2M_2\Big)+\frac{B_\mu}{8\pi^2}\Big(-\frac{3}{5}g_1^2-3g_2^2+\tr(y_{l}^2+3y_d^2+3y_u^2)\Big),\\
\frac{dm^2_{H_u}}{dt}=&\frac{1}{8\pi^2}\left(-\frac{3}{5}g_1^2M_1^2-3g^2_2M_2^2+\frac{3}{10}g_1^2 S+3\sum_{ij}y_{u,ij}^2(m^2_{H_u}+m^2_{\t q,i}+m^2_{\t u,j}) \right.\\
&\left.+ 2 \mu^2\left( -\frac{3}{5}g_1^2-3g_2^2+\tr(y_{l}^2+3y_d^2+3y_u^2)\right)
\right),\\
\frac{dm^2_{H_d}}{dt}=&\frac{1}{8\pi^2}\left(-\frac{3}{5}g_1^2M_1^2-3g^2_2M_2^2-\frac{3}{10}g_1^2 S+3\sum_{ij}y_{d,ij}^2(m^2_{H_d}+m^2_{\t q,i}+m^2_{\t d,j}) \right.\\
&\left.+\sum_{ij}y_{L,ij}^2(m^2_{H_d}+m^2_{\t l,i}+m^2_{\t e,j})+ 2 \mu^2\left( -\frac{3}{5}g_1^2-3g_2^2+\tr(y_{l}^2+3y_d^2+3y_u^2)\right)\right),
\end{align*}

\begin{align*}
\frac{dm^2_{\t q,i}}{dt}=&\frac{1}{8\pi^2}\Big(-\frac{1}{15}g_1^2M_1^2-3g^2_2M_2^2-\frac{16}{3}g_3^2 M_3^2+\frac{1}{10}g_1^2 S+\sum_{j}[y_{u,ij}^2(m^2_{H_u}+m^2_{\t q,i}+m^2_{\t u,j})\\
&+y_{d,ij}^2(m^2_{H_d}+m^2_{\t q,i}+m^2_{\t d,j})]\Big),\\
\frac{dm^2_{\t u,i}}{dt}=&\frac{1}{8\pi^2}\Big(-\frac{16}{15}g_1^2M_1^2-\frac{16}{3}g_3^2 M_3^2-\frac{2}{5}g_1^2 S+2\sum_{j}y_{u,ji}^2(m^2_{H_u}+m^2_{\t u,i}+m^2_{\t q,j})\Big),
\end{align*}
\begin{align*}
\frac{dm^2_{\t d,i}}{dt}=&\frac{1}{8\pi^2}\Big(-\frac{4}{15}g_1^2M_1^2-\frac{16}{3}g_3^2 M_3^2+\frac{1}{5}g_1^2 S+2\sum_{j}y_{d,ji}^2(m^2_{H_d}+m^2_{\t d,i}+m^2_{\t q,j})\Big),\;\;\;\;\;\;\;\;\\
\frac{dm^2_{\t l,i}}{dt}=&\frac{1}{8\pi^2}\Big(-\frac{3}{5}g_1^2M_1^2-3g^2_2M_2^2-\frac{3}{10}g_1^2 S+\sum_{j}y_{{l},ij}^2(m^2_{H_d}+m^2_{\t l,i}+m^2_{\t e,j})\Big),\\
\frac{dm^2_{\t e,i}}{dt}=&\frac{1}{8\pi^2}\Big(-\frac{12}{5}g_1^2M_1^2+\frac{3}{5}g_1^2 S+2\sum_{j}y_{{l},ji}^2(m^2_{H_d}+m^2_{\t e,i}+m^2_{\t l,j})\Big),
\end{align*}
where
\begin{align*}
 S=\tr(Ym^2)=m^2_{H_u}-m^2_{H_d}+\sum_i(m^2_{\t q,i}-2 m^2_{\t u,i}+m^2_{\t d,i}-m^2_{\t l,i}+m^2_{\t e,i}).
\end{align*}

The $\mu$-dependent terms in the $\beta$-functions of $m_{H_u}^2$ and $m_{H_d}^2$ arise because the $\mu$ term in our case corresponds to a higgsino mass, instead of a superpotential parameter. Note that in the scalar potential (\ref{eq:fullVH}, \ref{Vtop}) there are no $\mu$-terms appearing. 

Summarizing the approach described in \S \ref{sec:eft}, the RG flow to the scale of the top mass is implemented in a mass-dependent scheme by integrating out particles at their decoupling thresholds, which requires correcting the RG equations on each energy interval. For the soft masses, this amounts to dropping the contributions to the beta functions proportional to the soft masses of the integrated-out particles. 
In the case of the gauge couplings and gaugino masses (not shown above), the $\beta$-function coefficient is computed at each energy interval from the general one loop formula for a theory with an arbitrary number of scalar and fermion multiplets, only counting the fields with masses below the upper threshold.

To illustrate some of the points in \S \ref{subsec:comp-scale}, let us analyze the leading contributions to the RGEs of $H_u$ and stops,
\begin{eqnarray}\label{eq:beta-system}
8\pi^2\,\frac{dm^2_{H_u}}{dt}& \approx &3 y_{u,33}^2(m^2_{H_u}+m^2_{\t q,3}+m^2_{\t u,3}) -\frac{3}{5}g_1^2M_1^2-3g^2_2M_2^2+\frac{3}{10}g_1^2 S+3 y_{u,32}^2(m^2_{\t q,2}+m^2_{\t u,2})\nonumber\\
8\pi^2\,\frac{dm^2_{\t q,3}}{dt}&\approx& y_{u,33}^2(m^2_{H_u}+m^2_{\t q,3}+m^2_{\t u,3}) -\frac{1}{15}g_1^2M_1^2-3g^2_2M_2^2-\frac{16}{3}g_3^2 M_3^2+\nonumber\\
&+&\frac{1}{10}g_1^2 S+(y_{u,32}^2 m^2_{\t u,2}+y_{d,32}^2 m^2_{\t d,2})+y_{d,33}^2 m_{H_d}^2\nonumber\\
8\pi^2\,\frac{dm^2_{\t u,3}}{dt}&\approx&2 y_{u,33}^2(m^2_{H_u}+m^2_{\t q,3}+m^2_{\t u,3}) -\frac{16}{15}g_1^2M_1^2-\frac{16}{3}g_3^2 M_3^2-\frac{2}{5}g_1^2 S+y_{u,32}^2 m^2_{\t q,2}\,.
\end{eqnarray}
Contributions from inter-generational mixing with the heavy sfermions, the $U(1)_Y$ FI term and $H_d$ are now manifest. 

In the approximation where the heavy masses do not run, this closed system of equations can be easily diagonalized, yielding analytical expressions for $m_{Hu}^2$, $m_{\t q,3}^2$ and $m_{\t u,3}^2$ in terms of the microscopic parameters $(h \hat \mu, x, \lambda_u, \lambda_d)$. We find that composite fields give small contributions to the running of $m_{H_u}^2$ and $m_{\t q_3}^2$, the reason being a (partial) cancellation of the effects from $S$ and inter-generational mixing. The effect on $m_{\t u_3}^2$ is slightly larger, but still at the percent level.

On the other hand, there are strong effects on the running slepton mass, because the boundary soft mass is smaller, and contributions from $S$ and off-diagonal Yukawas now add up. This was analyzed in \S \ref{subsec:comp-scale}.

\subsection{Two loop effects}

In models with heavy first and/or second generations at the multi-TeV scale, it is known that two loop effects on the light third generation sfermions can become important and in fact dominate over one loop effects~\cite{ArkaniHamed:1997ab}. We now estimate such contributions in our context.

The two loop MSSM RGEs can be found in the second reference of~\cite{Martin:1993zk}. For example, the two loop contribution to the beta function for the third generation slepton is (excluding a factor of $1/(16 \pi^2)^2$)
\begin{equation}
\beta_{\t L_3}^{(2)} = \frac{621}{25} g_1^4 M_1^2+\frac{18}{5} g_2^2 g_1^2 \left(M_1^2+M_2 M_1+M_2^2\right)+33 g_2^4
  M_2^2-\frac{6 g_1^2 {\mathcal{S}'}}{5}+\frac{3}{5} g_1^2 \sigma _1+3 g_2^2 \sigma _2\,.
\end{equation}
Smaller contributions from Yukawas have been neglected, and
\begin{equation}
\begin{aligned}
{\mathcal{S}'} =& 2 \left(\frac{2 g_1^2}{15}+\frac{8 g_3^2}{3}\right) m_{{\t d_{1,2}}}^2+\frac{12}{5} g_1^2
  m_{{\t e_{1,2}}}^2+\left(\frac{3 g_1^2}{10}+\frac{3 g_2^2}{2}\right) \left(-m_{{H_d}}^2-2
  m_{{\t L_{1,2}}}^2\right) \cr
 &  +2 \left(\frac{g_1^2}{30}+\frac{3 g_2^2}{2}+\frac{8 g_3^2}{3}\right)
  m_{{\t q_{1,2}}}^2-2 \left(\frac{16 g_1^2}{15}+\frac{16 g_3^2}{3}\right) m_{{\t u_{1,2}}}^2\cr
\sigma_1 = &\frac{1}{5} g_1^2 \left(2 \left(2 m_{{\t d_{1,2}}}^2+6 m_{{\t e_{1,2}}}^2+3
  m_{{\t L_{1,2}}}^2+m_{{\t Q_{1,2}}}^2+8 m_{{u_{1,2}}}^2\right)+3 m_{{H_d}}^2\right) \cr
\sigma_2   =&   g_2^2 \left(m_{{H_d}}^2+2 \left(m_{{\t L_{1,2}}}^2+3 m_{{\t q_{1,2}}}^2\right)\right) \cr
\sigma_3 =& 2 g_3^2 \left(m_{{\t d_{1,2}}}^2+2 m_{{\t q_{1,2}}}^2+m_{{\t u_{1,2}}}^2\right) \,.
\end{aligned}
\end{equation}

Here we already made the simplifying assumptions appropriate in our context, that the first two generations give the dominant contribution and that, approximately their masses are degenerate. In particular $m_{\t u_{1,2}}$ is the average mass of the first and second generation.

Two loop contributions tend to decrease the soft masses.
Evaluating these effects in our range of parameters (where composites have mass $m_{CW} \sim 10 - 20$ TeV), the two loop correction to the stop squared mass is found to be approximately one order of magnitude smaller than the one loop effects we have included. These effects are also in general small for sleptons, except in the regime where the NLSP is the sneutrino. In this case they become important, and it would be interesting to understand better how they modify the properties of the NLSP. We leave this for future work.

\section{Comments about fine-tuning}\label{subsec:tuning}

In our proposal, supersymmetry breaking and EWSB have a unified origin, so it is interesting to understand how sensitive the vacuum $v_u \sim 174\,{\rm GeV}$ is to changes in the microscopic parameters (denoted below by `$a$'). The basic naturalness criterion, given for instance in~\cite{Barbieri:1987fn}, is that
\begin{equation}
\Delta(M_Z^2) \equiv \left|\frac{a}{M_Z^2}\,\frac{\partial M_Z^2}{\partial a} \right|
\end{equation}
should not be too large. (We refer the reader to~\cite{Essig:2007kh} for a bottom-up analysis of tuning over the MSSM parameter space, and for references.)

This places upper bounds on the masses of heavy superpartners, so the first worry is that heavy composite generations require un-naturally large cancellations. However, this is not the case because the Higgs mass is quite insensitive to quantum effects from such particles; this was found in~\cite{Cohen:1996vb, Dimopoulos:1995mi}, and can be seen directly from the smallness of the coefficient $c_{\t q,2}$ in (\ref{eq:analytic-mH}). This equation also shows that $m_{H_u}^2$ is quite insensitive to the heavy Higgs as well --in fact, there are partial cancellations between both effects. Composite masses of the order $10-20\,{\rm TeV}$ give a (mild) fine-tuning of order $10\%$.

The Higgs VEV is most sensitive to the masses of the stop and $H_u$. Naively, having a stop around $3\,{\rm TeV}$, as is the case in many of the examples above, leads to a fine tuning generally much larger than a part in 100. However, this low energy estimate is not completely correct, because the soft masses at the TeV scale are correlated, all being determined by a few microscopic parameters ($h\hat \mu\,,x\,,\lambda_u\,,\,h$). This tends to reduce the amount of fine-tuning. Furthermore, the solution (\ref{eq:analytic-mH}) reveals that the influence of the stop is somewhat smaller than expected. The total fine-tuning measure in the range of (\ref{eq:good-range}) is slightly below the percent level, so that the model is somewhat tuned. The underlying reason for this is the rather strong assumption that a single-sector dynamics be simultaneously responsible for the EW scale, soft parameters and flavor textures.


\bibliographystyle{JHEP}
\renewcommand{\refname}{Bibliography}
\addcontentsline{toc}{section}{Bibliography}
\providecommand{\href}[2]{#2}\begingroup\raggedright

\end{document}